%% file: cccp.tex
\documentclass[usenatbib]{mn2e}
\include{epsf}

\def \msun{\hbox{$\rm{M}_\odot$}}

\title[Weak lensing masses of CCCP clusters]{The Canadian Cluster
  Comparison Project: weak lensing masses and SZ scaling
  relations\thanks{Based on observations from the Canada-France-Hawaii
    Telescope, which is operated by the National Research Council of
    Canada, le Centre National de la Recherche Scientifique and the
    University of Hawaii.}}

\author[Hoekstra et al.]{Henk Hoekstra$^{1,2}$, Andisheh
  Mahdavi$^{3}$, Arif Babul$^2$ and Chris Bildfell$^2$\\
$^1$~Leiden Observatory, Leiden University, PO Box 9513, 2300 RA, Leiden, the
Netherlands\\ 
$^2$~University of Victoria, Dept. of Physics \& Astronomy, 3800 Finnerty Rd, 
Victoria, BC V8P 5C2, Canada\\
$^3$~San Francisco State University, Dept. of Physics \& Astronomy, 
1600 Holloway Avenue, San Francisco, CA 94132}

\begin{document}

\date{Accepted. Received; in original form}

\maketitle

\begin{abstract}
The Canadian Cluster Comparison Project is a comprehensive
multi-wavelength survey targeting 50 massive X-ray selected clusters
of galaxies to examine baryonic tracers of cluster mass and to probe
the cluster-to-cluster variation in the thermal properties of the hot
intracluster medium. In this paper we present the weak lensing masses,
based on the analysis of deep wide-field imaging data obtained using
the Canada-France-Hawaii-Telescope. The final sample includes two
additional clusters that were located in the field-of-view. We take
these masses as our reference for the comparison of cluster properties
at other wavelengths. In this paper we limit the comparison to
published measurements of the Sunyaev-Zel'dovich effect. We find
that this signal correlates well with the projected lensing mass,
with an intrinsic scatter of $12\pm5\%$ at $\sim r_{2500}$,
demonstrating it is an excellent proxy for cluster mass.

\end{abstract}


\section{Introduction}

Observational constraints on the various constituents that make up the
universe have tightened tremendously in the past decade. However, many
puzzles remain, most notably the origin of the accelerated expansion
of the universe \citep[e.g.,][]{Riess98, Perlmutter99}.  Another
challenge is to explain the transition from a smooth early universe,
as indicated by observations of the cosmic microwave background, to
today's highly structured and complex universe.

Arguably, clusters of galaxies provide one of the most important
pieces of this fascinating cosmic puzzle. Galaxy clusters are one of
the primary reservoirs of baryons in the local Universe that can be
studied using observations spanning the full electromagnetic
spectrum. As such, they provide an excellent laboratory for studying
the physics affecting the evolution of baryons, at least over the past
10 Gyrs. Finally, the number density of clusters is a sensitive
function of key cosmological parameters, including the dark energy
equation of state \citep[see e.g.,][for a recent review]{Allen11}. Hence,
there is an active ongoing effort to use clusters as precision probes
of these parameters.

The use of clusters as cosmological probes and the study of the baryon
physics to explain their observable properties are closely
connected. The former requires cluster catalogs with well-defined
selection functions that somehow need to be related to predictions.
The predictions are based on numerical simulations of cold dark
matter. As clusters are discovered through their optical, X-ray or
radio properties, it is imperative that we understand the relation
between the observables and the underlying dark matter distribution.

Cluster samples are increasing rapidly thanks to optical surveys
\citep[e.g.,][]{Gladders05,Koester07}, X-ray studies
\citep[e.g.,][]{Reiprich02,Ebeling10} and Sunyaev-Zel'dovich surveys
\citep[e.g.,][]{Williamson11,Marriage11}. The detection significance
correlates with mass and can thus be used as a mass-proxy, which
nonetheless needs to be calibrated. This can be done using deep X-ray
observations, under the assumption that the intra-cluster medium (ICM)
is in hydrostatic equilibrium. There is, however, both theoretical and
observational evidence that that X-ray masses tend to be biased low if
one assumes hydrostatic equilibrium \citep[e.g.,][]{Nagai07,Mahdavi08,
  Mahdavi12}.

To compare the baryonic properties of clusters to the results from
numerical simulations, we instead need a direct probe of the (dark)
matter distribution. Such a probe exists in the form of weak
gravitational lensing: the gravitational potential of the cluster
perturbs the paths of photons emitted by distant galaxies. As a
result, the images of the galaxies appear slightly distorted. The
amplitude of the distortion provides us with a direct measurement of
the gravitational tidal field, which in turn can be used to map the
distribution of dark matter {\it along the line-of-sight} directly
\citep[e.g.,][]{KS93}.  However, comparison with other mass traces
does require one to make assumptions about the geometry of the
cluster, because weak lensing measures the total mass projected along
the line-of-sight.

Weak gravitational lensing is now a well-established technique to
study the distribution of matter in the universe. The applications
range from the study of galaxy halos
\citep[e.g.,][]{Hoekstra04,Mandelbaum06, Uitert11} to the study of
large-scale structure \citep[e.g.,][]{Hoekstra06,Fu08,Schrabback10}.
Several developments in the past decade have also led to improvements
in the weak lensing studies of galaxy clusters. For instance, the
galaxy shapes are not only affected by gravitational lensing, but
observational distortions can cause systematic signals that are
similar in size.  However, extensive tests have shown that techniques can
now reach an accuracy of $1-2\%$
\citep[e.g.,][]{STEP1,STEP2,GREAT08,GREAT10}.  Finally, the source
redshift distributions, which are required to convert the lensing
signal into a physical mass, are now much better known, compared to
even a few years ago.

It is possible to derive mass estimates by fitting parametric models
to the data, which is the only option if the data extend to small
radii \citep[e.g][]{Smail97, Dahle02, Cypriano04, Jee11}. This is also
relevant for the study of high redshift $(z>0.6)$ clusters, which can
only be studied reliably using Hubble Space Telescope observations.  At
these high redshifts lensing masses are particularly important,
because the clusters are expected to be dynamically young. Fortunately
the number of high redshift clusters for which weak lensing masses
have been determined has increased in recent year. For instance
\cite{Jee11} analysed a sample of 22 $z>0.9$ clusters and found
evidence for an evolution in the normalization of the relation between
mass and $T_X$, the X-ray temperature. Clusters of galaxies, however,
may show significant substructure or are not well described by the
adopted model, which leads to biases in the mass estimates
\citep{Hoekstra02}. As a consequence, it is important to measure the
lensing signal out to large radii, which allows for more direct mass
measurements.

This is now possible thanks to wide-field imagers on world-class
telescopes. For instance, \cite{Hoekstra07} presented masses for a
sample of 20 X-ray luminous clusters of galaxies that were observed
using the CFH12k camera on the Canada-France-Hawaii Telescope (CFHT).
As part of the Local Cluster Substructure Survey (LoCuSS),
\cite{Okabe10} presented results for a sample of 30 clusters with
$0.15<z<0.3$ using Subaru data. As part of a weak lensing follow-up
program of clusters discovered by the South Pole Telescope,
\cite{High12} presented results for an initial sample of 5 clusters
observed with the Megacam imager on the Magellan telescope.

To fully exploit the statistical power of cluster surveys, it is
timely to increase the sample of clusters for which accurate weak
lensing masses are available. In this paper we update the results
presented in \cite{Hoekstra07} and augment the sample with 30 clusters
with redshifts $0.15<z<0.55$ that were observed using MegaCam on CFHT.
This sample forms the basis for the Canadian Cluster Comparison
Project (CCCP), which is a comprehensive multi-wavelength study of
these massive clusters of galaxies.

The structure of the paper is as follows. In \S2 we present the data
and discuss the data analysis. The weak lensing analysis is discussed
in \S3 and the mass measurements are presented in \S4. In \S5 we
compare our results to measurements of the Sunyaev-Zel'dovich Effect.
Throughout the paper we assume a cosmology with $\Omega_m=0.3$,
$\Omega_\Lambda=0.7$ and $H_0=70 h_{70}$~km/s/Mpc.

\section{Cluster Sample and Optical Data}

The main objective of the Canadian Cluster Comparison Project (CCCP)
is to study the different baryonic tracers of cluster mass and to
explore insights about the thermal properties of the hot diffuse gas
and the dynamical states of the clusters that can be gained from
cluster-to-cluster variations in these relationships. An important
aspect is to compare to accurate weak lensing masses for the clusters,
which requires deep data with good image quality over a wide
field-of-view, for which we use the Canada-France-Hawaii
Telescope (CFHT). 

\begin{figure}
\begin{center}
\leavevmode \hbox{%
\epsfxsize=8.5cm \epsffile{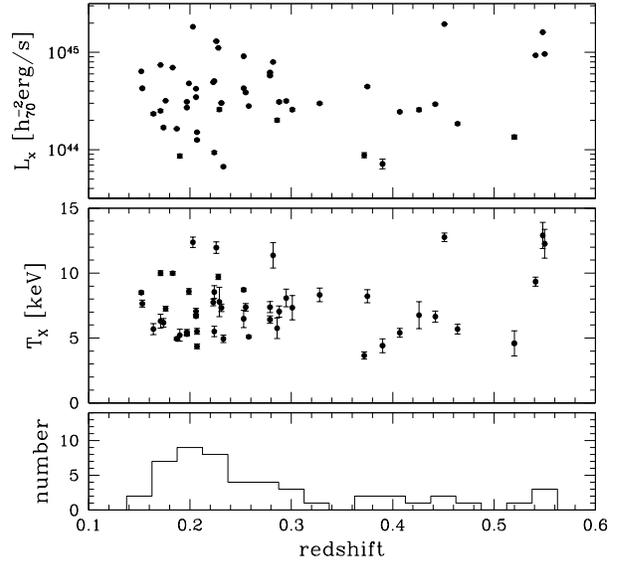}}
\caption{X-ray luminosity within $r_{500}$ in the $2-10$ keV band
  (top) and temperature within $r_{2500}$ (middle) as a function of
  cluster redshift. The distribution of redshifts is shown in the
  bottom panel.\label{fig:sample}}
\end{center}
\end{figure}

The starting point of the project is the sample of 20 clusters, for
which deep archival $B$ and $R$ band observations with the CFH12k
camera were available (the first 20 entries in in
Table~\ref{tabsample}). This sample was studied in
\cite{Hoekstra07,Mahdavi08}. Nearly half of these clusters were
originally observed by the Canadian Network for Observational
Cosmology\citep[CNOC1;][]{CNOC1, Carlberg96} and comprise the
brightest clusters in the {\it Einstein Observatory} Extended Medium
Sensitivity Survey \citep[EMSS;][]{Gioia90}.

To improve the statistics, this initial sample was augmented with an
additional 30 clusters (listed in Table~\ref{tabsample}) that were
observed using Megacam on CFHT in the $g'$ and $r'$-band. To ensure a
significant detection of the lensing signal, clusters with an ASCA
temperature of $k_B T_X>5$ keV, and redshifts $0.15<z<0.55$ were
selected based on the results from \cite{Horner01}. Note that 4 of the
EMSS clusters studied by \cite{Hoekstra07} do not meet the X-ray
temperature criterion\footnote{These are MS1224.7+2007, MS1231.3+1542,
  MS1455.0+2232 and MS1512.4+3647.}. Finally Table~\ref{tabsample}
lists two additional clusters (Abell~222 and Abell~1234) that were
located in the observed fields. 

In this paper we therefore present weak lensing masses for 52 clusters
of galaxies. The imaging data have been used in a number of related
studies. For instance, the surface brightness profiles of the
brightest cluster galaxies were studied in \cite{Bildfell08}. The CCCP
sample also provided the high redshift subset in a study of the
evolution of the dwarf-to-giant ratio by \cite{Bildfell12}.

The X-ray analysis is described in detail in \cite{Mahdavi12} and
the resulting X-ray luminosity and temperature as a function of
redshift are presented in Figure~\ref{fig:sample} (top and middle
panel, resp.), as well as the distribution of cluster redshifts
(bottom panel). The sample is dominated by clusters at $z\sim 0.2$,
with a tail out to $z\sim 0.5$. This mostly reflects the limitations
of the input ASCA catalog. It is important to note, however, that no
selection was made based on the optical properties or the dynamical
state of the cluster. Interestingly, although the sample is by no
means complete, \cite{Mahdavi12} find that the X-ray properties appear
to be representative. The statistics of the $L_X-T_X$ relation of the
CCCP sample are indistinguishable from samples with well-characterized
selection functions, such as MACS \citep{Ebeling10} and HIFLUGCS
\citep{Reiprich02}.

\begin{table*}
\begin{center}
\caption{Basic information of the Canadian Cluster Comparison Project sample\label{tabsample}}
\begin{tabular}{lllllllll}
\hline
\hline
(1) & (2) & (3) & (4) & (5) & (6) & (7) & (8) & (9)\\ 
  & name      & $z$ & RA & DEC & offset & mag & $\langle\beta\rangle$ & $\langle\beta^2\rangle$ \\
  &            &     & (J2000.0) & (J2000.0) & [$h_{70}^{-1}$kpc] & & & \\
\hline
1  & Abell 68       & 0.255 & $00^{\rm h}37^{\rm m}06.9^{\rm s}$ & $+09^\circ09'24''$ & 12  & 21-25   & 0.50 & 0.26 \\
2  & Abell 209      & 0.206 & $01^{\rm h}31^{\rm m}52.5^{\rm s}$ & $-13^\circ36'40''$ & 49  & 21-25   & 0.63 & 0.39 \\
3  & Abell 267      & 0.230 & $01^{\rm h}52^{\rm m}42.0^{\rm s}$ & $+01^\circ00'26''$ & 85  & 20-25   & 0.52 & 0.28 \\
4  & Abell 370      & 0.375 & $02^{\rm h}39^{\rm m}52.7^{\rm s}$ & $-01^\circ34'18''$ & 93  & 22-25   & 0.41 & 0.17 \\
5  & Abell 383      & 0.187 & $02^{\rm h}48^{\rm m}03.4^{\rm s}$ & $-03^\circ31'44''$ & 4   & 21-24.5 & 0.64 & 0.41 \\
6  & Abell 963      & 0.206 & $10^{\rm h}17^{\rm m}03.8^{\rm s}$ & $+39^\circ02'51''$ & 5   & 21-25   & 0.59 & 0.35 \\
7  & Abell 1689     & 0.183 & $13^{\rm h}11^{\rm m}30.0^{\rm s}$ & $-01^\circ20'30''$ & 6   & 21-24.5 & 0.64 & 0.41 \\
8  & Abell 1763     & 0.223 & $13^{\rm h}35^{\rm m}20.1^{\rm s}$ & $+41^\circ00'04''$ & 84  & 21-25   & 0.57 & 0.33 \\
9  & Abell 2218     & 0.176 & $16^{\rm h}35^{\rm m}48.8^{\rm s}$ & $+66^\circ12'51''$ & 36  & 21-24.5 & 0.64 & 0.42 \\
10 & Abell 2219     & 0.226 & $16^{\rm h}40^{\rm m}19.9^{\rm s}$ & $+46^\circ42'41''$ & 26  & 21-25   & 0.56 & 0.32 \\
11 & Abell 2390     & 0.228 & $21^{\rm h}53^{\rm m}36.8^{\rm s}$ & $+17^\circ41'44''$ & 2   & 21-25   & 0.59 & 0.34 \\
12 & MS 0015.9+1609 & 0.547 & $00^{\rm h}18^{\rm m}33.5^{\rm s}$ & $+16^\circ26'16''$ & 50  & 22-25.5 & 0.27 & 0.07 \\
13 & MS 0906.5+1110 & 0.170 & $09^{\rm h}09^{\rm m}12.6^{\rm s}$ & $+10^\circ58'28''$ & 2   & 21-25   & 0.67 & 0.46 \\
14 & MS 1224.7+2007 & 0.326 & $12^{\rm h}27^{\rm m}13.5^{\rm s}$ & $+19^\circ50'56''$ & $-$ & 21-25   & 0.44 & 0.20 \\
15 & MS 1231.3+1542 & 0.235 & $12^{\rm h}33^{\rm m}55.4^{\rm s}$ & $+15^\circ25'58''$ & 22  & 21-25.5 & 0.58 & 0.34 \\
16 & MS 1358.4+6245 & 0.329 & $13^{\rm h}59^{\rm m}50.6^{\rm s}$ & $+62^\circ31'05''$ & 2   & 21-25   & 0.45 & 0.20 \\
17 & MS 1455.0+2232 & 0.257 & $14^{\rm h}57^{\rm m}15.1^{\rm s}$ & $+22^\circ20'35''$ & 7   & 21-25.5 & 0.56 & 0.32 \\
18 & MS 1512.4+3647 & 0.373 & $15^{\rm h}14^{\rm m}22.5^{\rm s}$ & $+36^\circ36'21''$ & 3   & 21-25.5 & 0.43 & 0.19 \\
19 & MS 1621.5+2640 & 0.428 & $16^{\rm h}23^{\rm m}35.5^{\rm s}$ & $+26^\circ34'14''$ & 54  & 21-25.5 & 0.37 & 0.14 \\
20 & CL0024.0+1652  & 0.390 & $00^{\rm h}26^{\rm m}35.6^{\rm s}$ & $+17^\circ09'44''$ & 25  & 22-25.5 & 0.38 & 0.14 \\
\hline
21 & Abell 115N   & 0.197 & 00$^{\rm h}$55$^{\rm m}$50.6$^{\rm s}$ & $+26^\circ24'38''$ & 11      & 20-25 & 0.61 & 0.38 \\
   & Abell 115S   & 0.197 & 00$^{\rm h}$56$^{\rm m}$00.3$^{\rm s}$ & $+26^\circ20'33''$ & 9       & 20-25 & 0.61 & 0.38 \\
22 & Abell 222    & 0.213 & 01$^{\rm h}$37$^{\rm m}$34.0$^{\rm s}$ & $-12^\circ59'29''$ & 14      & 20-25 & 0.58 & 0.35 \\
23 & Abell 223N   & 0.207 & 01$^{\rm h}$38$^{\rm m}$02.3$^{\rm s}$ & $-12^\circ45'20''$ & $-$     & 20-25 & 0.59 & 0.36 \\
   & Abell 223S   & 0.207 & 01$^{\rm h}$37$^{\rm m}$56.0$^{\rm s}$ & $-12^\circ49'10''$ & 10      & 20-25 & 0.59 & 0.36 \\
24 & Abell 520    & 0.199 & 04$^{\rm h}$54$^{\rm m}$10.1$^{\rm s}$ & $+02^\circ55'18''$ & 430$^a$ & 20-25 & 0.61 & 0.38 \\
25 & Abell 521    & 0.253 & 04$^{\rm h}$54$^{\rm m}$06.9$^{\rm s}$ & $-10^\circ13'25''$ & 47      & 21-25 & 0.53 & 0.29 \\
26 & Abell 586    & 0.171 & 07$^{\rm h}$32$^{\rm m}$20.3$^{\rm s}$ & $+31^\circ38'01''$ & 14      & 20-25 & 0.63 & 0.40 \\
27 & Abell 611    & 0.288 & 08$^{\rm h}$00$^{\rm m}$56 8$^{\rm s}$ & $+36^\circ03'24''$ & 12      & 21-25 & 0.48 & 0.24 \\
28 & Abell 697    & 0.282 & 08$^{\rm h}$42$^{\rm m}$57.6$^{\rm s}$ & $+36^\circ21'59''$ & 20      & 21-25 & 0.50 & 0.26 \\
29 & Abell 851    & 0.407 & 09$^{\rm h}$42$^{\rm m}$57.5$^{\rm s}$ & $+46^\circ58'50''$ & 231$^a$ & 22-25 & 0.38 & 0.15 \\
30 & Abell 959    & 0.286 & 10$^{\rm h}$17$^{\rm m}$36.0$^{\rm s}$ & $+59^\circ34'02''$ & 39      & 21-25 & 0.50 & 0.26 \\
31 & Abell 1234   & 0.166 & 11$^{\rm h}$22$^{\rm m}$30.0$^{\rm s}$ & $+21^\circ24'22''$ & $-$     & 20-25 & 0.66 & 0.45 \\
32 & Abell 1246   & 0.190 & 11$^{\rm h}$23$^{\rm m}$58.8$^{\rm s}$ & $+21^\circ28'50''$ & $-$     & 20-25 & 0.62 & 0.40 \\
33 & Abell 1758E  & 0.279 & 13$^{\rm h}$32$^{\rm m}$38.4$^{\rm s}$ & $+50^\circ33'36''$ & 318     & 21-25 & 0.52 & 0.27 \\
   & Abell 1758W  & 0.279 & 13$^{\rm h}$32$^{\rm m}$52.1$^{\rm s}$ & $+50^\circ31'34''$ & 56      & 21-25 & 0.52 & 0.27 \\
34 & Abell 1835   & 0.253 & 14$^{\rm h}$01$^{\rm m}$02.1$^{\rm s}$ & $+02^\circ52'43''$ & 12      & 21-25 & 0.53 & 0.29 \\
35 & Abell 1914   & 0.171 & 14$^{\rm h}$26$^{\rm m}$02.8$^{\rm s}$ & $+37^\circ49'28''$ & 227$^a$ & 20-25 & 0.66 & 0.44 \\
36 & Abell 1942   & 0.224 & 14$^{\rm h}$38$^{\rm m}$21.9$^{\rm s}$ & $+03^\circ40'13''$ & 0.7     & 20-25 & 0.58 & 0.34 \\
37 & Abell 2104   & 0.153 & 15$^{\rm h}$40$^{\rm m}$07.9$^{\rm s}$ & $-03^\circ18'16''$ & 10      & 20-25 & 0.68 & 0.47 \\
38 & Abell 2111   & 0.229 & 15$^{\rm h}$39$^{\rm m}$40.5$^{\rm s}$ & $+34^\circ25'27''$ & 108     & 20-25 & 0.57 & 0.33 \\
39 & Abell 2163   & 0.203 & 16$^{\rm h}$15$^{\rm m}$49.0$^{\rm s}$ & $-06^\circ08'41''$ & 11      & 20-25 & 0.60 & 0.36 \\
40 & Abell 2204   & 0.152 & 16$^{\rm h}$32$^{\rm m}$47.0$^{\rm s}$ & $+05^\circ34'33''$ & 4       & 20-25 & 0.68 & 0.46 \\
41 & Abell 2259   & 0.164 & 17$^{\rm h}$20$^{\rm m}$09.7$^{\rm s}$ & $+27^\circ40'08''$ & 75      & 20-25 & 0.65 & 0.44 \\
42 & Abell 2261   & 0.224 & 17$^{\rm h}$22$^{\rm m}$27.2$^{\rm s}$ & $+32^\circ07'58''$ & 5       & 20-25 & 0.57 & 0.33 \\
43 & Abell 2537   & 0.295 & 23$^{\rm h}$08$^{\rm m}$22.2$^{\rm s}$ & $-02^\circ11'32''$ & 8       & 21-25 & 0.49 & 0.24 \\
44 & MS0440.5+0204     & 0.190 & 04$^{\rm h}$43$^{\rm m}$09.9$^{\rm s}$ & $+02^\circ10'19''$      & 4 & 20-25 & 0.60 & 0.37 \\
45 & MS0451.6-0305     & 0.550 & 04$^{\rm h}$54$^{\rm m}$10.8$^{\rm s}$ & $-03^\circ00'51''$      & 34 & 22-25 & 0.28 & 0.08 \\
46 & MS1008.1-1224     & 0.301 & 10$^{\rm h}$10$^{\rm m}$32.3$^{\rm s}$ & $-12^\circ39'53''$      & 12 & 21-25 & 0.47 & 0.23 \\
47 & RXJ1347.5-1145    & 0.451 & 13$^{\rm h}$47$^{\rm m}$30.1$^{\rm s}$ & $-11^\circ45'09''$      & 7 & 22-25 & 0.34 & 0.12 \\
48 & RXJ1524.6+0957    & 0.516 & 15$^{\rm h}$24$^{\rm m}$41.6$^{\rm s}$ & $+09^\circ59'34''$      & 44 & 22-25 & 0.30 & 0.09 \\
49 & MACS J0717.5+3745 & 0.548 & 07$^{\rm h}$17$^{\rm m}$30.4$^{\rm s}$ & $+37^\circ45'38''$      & 113 & 22-25 & 0.27 & 0.07 \\
50 & MACS J0913.7+4056 & 0.442 & 09$^{\rm h}$13$^{\rm m}$45.5$^{\rm s}$ & $+40^\circ56'29''$      & 1 & 22-25 & 0.36 & 0.13 \\
51 & CIZA J1938+54     & 0.260 & 19$^{\rm h}$38$^{\rm m}$18.1$^{\rm s}$ & $+54^\circ09'40''$      & $-$ & 21-25 & 0.52 & 0.27 \\
52 & 3C295             & 0.460 & 14$^{\rm h}$11$^{\rm m}$20.6$^{\rm s}$ & $+52^\circ12'10''$      & 5 & 22-25 & 0.34 & 0.12 \\
\hline
\hline
\end{tabular}
\bigskip
\begin{minipage}{0.9\linewidth}
{\footnotesize Column 2: cluster name; Column 3: cluster redshift;
  Column 4,5: right ascension and declination (J2000.0) of the adopted
  cluster center. In all but three cases (Abell~520, Abell~851 and
  Abell 1914) we take this to be the position of the brightest cluster
  galaxy (BCG). Column 6: the offset between the peak in the X-ray
  emission and the position of the BCG. For the values marked $^a$ we
  actually use the X-ray position as the cluster center. Note that not
  all clusters have modern X-ray data; Column 7: magnitude range used
  for the source galaxies. For clusters $1-20$ this is the $R_C$
  filter and $r'$ for the remaining clusters; Column 8,9: the average
  values of $\beta=D_{ls}/D_{s}$ and $\langle\beta^2\rangle$ (as
  explained in the text)}
\end{minipage}

\end{center}
\end{table*}

For the clusters observed with Megacam and $z<0.3$ the observations
typically consist of four 400s exposures in $g'$ and eight 600s
exposures in $r'$. For the higher redshift clusters we obtained four
600s exposure in $g'$ and twelve 600s exposures in $r'$. These
integrations times enable us to use galaxies down to $r'=25$ in our
weak lensing analysis. We only use the $r'$ data because they are
deeper and have better image quality.  The $g'$ data are used to
identify the cluster early-type galaxies, which lie on a well-defined
color-magnitude relation. The exposure times and information about the
data reduction of the clusters observed using the CFH12k camera in the
$B$ and $R$ filters can be found in \cite{Hoekstra07}.

\subsection{Processing of optical imaging data}

Current wide field imaging instruments such as MegaCam consist of a
mosaic of chips, and special care needs to be taken to account for
sudden jumps in the PSF properties when data from different exposures
are combined. To avoid such problems altogether, the CCCP data are
obtained in two sets of exposures.  Each set of four (or six)
exposures is taken with small dithers. The two sets are offset by
approximately half a chip in each direction to fill in most of the
gaps between chips. Rather than combining all our data into a single
deep frame, for our weak lensing analysis we consider each set
separately and combine the measurements in the catalog stage.

The data were detrended using the Elixir pipeline developed at
CFHT. The pipeline also provides photometric zeropoints and in most
cases a reasonable first order astrometric solution. We verified that
the data were indeed taken during photometric conditions. The
astrometric solution provided by Elixir is not sufficiently accurate
to combine the exposures into deeper images without affecting the PSF:
errors in the astrometry lead to additional anisotropies in the
images. Instead, we follow the procedure described in
\cite{Hoekstra06} to refine the astrometric solution. We use the
USNO-A2 catalog to calibrate the red images from the Digital Sky
Survey (POSS~II), which in turn is used to generate a catalog of
sources with accurate astrometric positions.

This new astrometric catalog is matched to each of the MegaCam images.
Each set of exposures is then processed separately. We detect objects
in each of the exposures and combine matched objects into a master
catalog, which contains the average positions of the matched
objects. This master catalog is used to derive the final second order
astrometric solution for each chip. This step ensures that the objects
in each set of exposures are accurately matched to the same position.

\section{Weak lensing analysis}

We briefly review the steps in the weak lensing analysis, but refer
the reader to \cite{Hoekstra07} for a detailed discussion of the
analysis and various issues that arise when interpreting the data.
The first step is to detect the faint galaxies and identify the
stars. The stars are used to correct the galaxy shapes for the effects
of PSF anisotropy and seeing. To do so, we use the procedure developed
by \cite{KSB95} and \cite{LK97}.  Modifications to the original
approach are described in \cite{Hoekstra98} and \cite{Hoekstra00}.
The same analysis pipeline was also used for the cosmic shear analysis
presented in \cite{Hoekstra06}.  Furthermore, the method also
performed well in the Shear Testing Programme \citep{STEP1,STEP2},
which demonstrated that we can recover the weak lensing signal with an
accuracy of $\sim 2\%$.

As mentioned earlier, each set of stacked images is analysed
separately.  Because the images were taken with small offsets, we can
analyse the data on a chip-by-chip basis. We thus obtain two catalogs
for each cluster of galaxies. The two catalogs are merged and the
measurements for the objects that appear in both are combined. The
resulting catalog is used to derive the cluster masses presented
below. However, the redundancy in the data allow us also to check for
consistency of the lensing signal.

To quantify the lensing signal we consider the azimuthally averaged
tangential shear $\langle\gamma_T\rangle$ as a function of distance
$r$ from the cluster centre (for which we use the positions listed in
Table~\ref{tabsample}; also see \S\ref{sec:center}). To minimize the
effect of the gaps between the chips we only consider the signal
between $0.5-2h_{70}^{-1}$Mpc and fit a singular isothermal sphere
model:

\begin{equation}
\langle\gamma_T\rangle=\frac{r_E}{2r},
\end{equation}

\noindent where $r_E$ is the Einstein radius. Under the assumption of
isotropic orbits and spherical symmetry, the Einstein radius (in
radians) is related to the line-of-sight velocity dispersion $\sigma$ through

\begin{equation}
r_E=4\pi\left(\frac{\sigma}{c}\right)^2\langle\beta\rangle,
\end{equation}

\noindent where $\langle\beta\rangle=\langle D_{ls}/D_{s}\rangle$ is
the average ratio of the angular diameter distances between the lens
and the source, and the observer and the source. The value of
$\langle\beta\rangle$ depends on the source redshift distribution, and
is discussed in detail in the following section.  For our comparison
we do need to account for small variations in the actual source
redshift distribution because the two images may have different
seeing. This is particularly true for pointings that were observed a
year apart. We therefore follow the procedure described below and also
determine average values for $\beta$ for each pointing.

\begin{figure}
\begin{center}
\leavevmode \hbox{%
\epsfxsize=8.5cm \epsffile{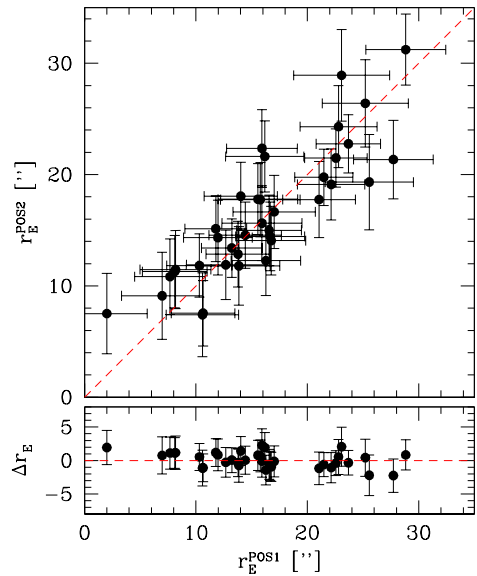}}
\caption{{\it top panel:} Plot of the best fit Einstein radii for the
  two sets of exposures. We have accounted for the small differences
  in the value for $\beta$ for the two sets. {\it bottom panel:}
  Residuals from the one-to-one relation between Einstein radii. The
  lensing signals inferred from both data sets agree very well and the
  observed scatter agrees well with the expected variation.
\label{comp_re}}
\end{center}
\end{figure}

Figure~\ref{comp_re} shows the results of our comparison: the
independent measurements of the lensing signal for each cluster agree
very well. For the difference in $r_E$ between the two positions we
find $\langle\Delta r_E\rangle=0.07\pm0.37$. The reduced
$\chi^2=0.51$, which is too small if the two measurements would have
been uncorrelated. However, the uncertainty in $r_E$ is determined by
the error in the shape measurements, which is a combination of the
noise in the image and the intrinsic shapes of the source
galaxies. The latter will cause the measurements to be correlated and
we need to account for this in the comparison.  We find that due to
gaps between the chips, masks, etc. $\sim 70\%$ of the objects appear
in both catalogs within the aperture we use to fit the model. We
estimate the expected variation in the difference in $r_E$ by
simulating the effect of image noise and shape noise from objects that
appear in only one of the catalogs. We find that the random error is
approximately $70\%$ of the total (formal) error and we expect a reduced
$\chi^2\sim0.5$, which is in good agreement with the actual value.
This suggests that the results obtained for the two positions are
indeed in good agreement with one another.

\subsection{Choice of cluster center}\label{sec:center}

To quantify the lensing signal we need to define the cluster center.
If the adopted center is offset from the `true' center the tangential
shear is lowered and the inferred mass is also biased low. Hence, one
option is to determine the location that maximizes the lensing signal,
for instance by reconstructing the projected mass distribution
\citep[e.g.,][]{KS93}, or fitting a model to the data. Such an
approach, however, will lead to masses that are biased high. Instead
we (typically) choose the location of the brightest cluster galaxy
(BCG) to define the cluster center.

The adopted centers are listed in Table~\ref{tabsample}. The projected
offset between the location of the BCG and the peak of the X-ray
emission determined by \cite{Mahdavi12} are also listed. The
distribution of offsets is presented in Figure~\ref{bias_offset}. With
the exception of the four clusters discussed below, the offsets are
less than $~\sim 100$kpc, resulting in negligible biases in the
cluster masses (see Figure~\ref{bias_offset} and discussion in
\S\ref{sec:apmass}). In three cases we use the position of the peak of
the X-ray emission. Abell~520 is a well known merging system, and
possibly the result of a three-way merger \citep{Mahdavi07, Okabe08,
  Jee12}. The unrelaxed cluster Abell~851 lacks a well defined
brightest cluster galaxy and the X-ray emission peaks away from the
concentration of bright cluster members. The BCG in Abell~1914 also
shows a large offset from the peak in the X-ray emission. It is
believed to be an ongoing merger \citep{Dahle02, Jones05, Okabe08},
showing high brightness tidal features \citep{Feldmeier04} and a radio
halo \citep{Giovannini99,Kempner01}.  Abell~1758 is also a merging
system with two peaks in the X-ray emission \citep{David04,
  Ragozzine12}. One peak is well centred on one of the BCGs, whereas
the other component is clearly offset from the other BCG, which we
nonetheless adopt as the center of the other cluster.

\subsection{Source redshift distribution}\label{sec:zdist}

To relate the lensing signal to physical quantities, such as mass,
requires knowledge of the redshifts of the faint source galaxies. The
higher the source redshift, the higher the lensing signal for a given
mass. The redshifts need not be known with high precision, and
photometric redshifts are sufficient. We lack the color information to
derive photometric redshifts for our sources, but fortunately it
suffices to assume an average redshift distribution that can be
obtained from other data sets. This does lead to an increase in the
uncertainty with which the cluster mass can be determined. As shown in
\cite{Hoekstra11b} this contribution is small for clusters with
$z\sim0.2$, and only becomes relevant when studying high redshift
clusters ($z>0.6$).

We therefore can use the photometric redshifts from \cite{Ilbert06}
which are based on the CFHT Legacy Survey Deep fields and well matched
in depth to what we need. Importantly, the survey covers a
sufficiently large area of sky (4 fields of one square degree each)
that sample variance, which has plagued earlier studies, is no longer
important.

The use of photometric redshifts naturally accounts for galaxies in
front of the cluster, which do not contribute to the lensing
signal. However, the regions around clusters show large enhancements
of galaxies (after all, that is what clusters of galaxies are) and
many of those will end up in our catalog of sources. Our data only
allow us to remove galaxies on the red-sequence. For our choice of
filters, this only leads to a modest reduction in the level of
contamination, because most faint cluster members are in fact blue.

The level of contamination after removing galaxies on the red-sequence
depends on the choice of filter. For instance, \cite{Okabe10} find
that their $V$ and $i'$ data allows for a relatively clean selection
of sources. As discussed below, compared to our Megacam observations,
the contamination is also lower for the clusters studied in
\cite{Hoekstra07}, which are based on $B$ and $R_C$ data. Hence our
combination of $g'$ and $r'$ is not ideal. As shown in \cite{High12},
an efficient way to remove cluster members for our sample would be to
obtain additional $i'$ data. We lack such data and therefore we follow
\cite{Hoekstra07} and derive an average correction for each cluster,
by measuring the excess of source galaxies as a function of radius.
We assume that the excess declines as $f_{cg}(r)\propto r^{-1}$, which
is a reasonable description of the observed excess counts
\citep[c.f. Figure~3 in][]{Hoekstra07}. Note that we do exclude the
galaxies on the red-sequence. To correct for the contamination by
cluster members, the observed tangential shear is then scaled by a
factor $1+f_{cg}(r)$.

The correction leads to a $\sim 13\%$ increase in the best fit
Einstein radii for the Megacam data, but only a $\sim 4\%$ increase
for the CFH12k data.  For the aperture masses the correction depends
on the overdensity. We find that on average the results for $M_{2500}$
are boosted by $\sim 18$ and $\sim 6\%$ for the Megacam and CFH12k
data, respectively. The contamination is less important for $M_{500}$,
which is based on measurements at larger radii: the masses based on
the Megacam data are on average increased by $\sim 8\%$, whereas the
masses of the other clusters are boosted by only $\sim 3\%$.

\begin{table*}
\begin{center}
\caption{Canadian Cluster Comparison Project sample\label{tabmass}}
\begin{tabular}{llcccccccccc}
\hline
(1) & (2) & (3) & (4) & (5) & (6) & (7) & (8) & (9) & (10) & (11) \\ 
name           & $r_E$ & $\sigma$ & $M^{\rm proj}_{0.5}$ & $r_{2500}$ & $M^{\rm ap}_{2500}$ & $r_{500}$ & $M^{\rm ap}_{500}$ & $M_{\rm vir}^{\rm NFW}$ & $M_{2500}^{\rm NFW}$ & $M_{500}^{\rm NFW}$ \\   
               & [arcsec]  & [km/s] &                       & [$h_{70}^{-1}$kpc] & & [$h_{70}^{-1}$kpc]  \\  
\hline
Abell 68       & $17.5\pm2.6$ & $1098^{+74}_{-80}$ & $4.0\pm0.6$ & $520^{+32}_{-33}$ & $2.6^{+0.6}_{-0.5}$ & $1178^{+95}_{-87}$ & $6.0^{+1.9}_{-1.6}$ & $13.6^{+3.4}_{-3.0}$ & $2.7^{+0.7}_{-0.6}$ & $7.5^{+1.9}_{-1.7}$\\
Abell 209      & $15.9\pm2.8$ & $937^{+76}_{-82}$  & $3.2\pm0.5$ & $477^{+26}_{-31}$ & $1.9^{+0.4}_{-0.4}$ & $1254^{+65}_{-69}$ & $6.9^{+1.5}_{-1.5}$ & $8.9^{+2.4}_{-2.2}$ & $1.8^{+0.5}_{-0.5}$ & $5.0^{+1.3}_{-1.2}$\\
Abell 267      & $14.8\pm2.8$ & $990^{+86}_{-94}$ & $3.5\pm0.6$  & $489^{+26}_{-26}$ & $2.1^{+0.4}_{-0.4}$ & $1144^{+94}_{-89}$ & $5.4^{+1.7}_{-1.5}$ & $9.7^{+2.8}_{-2.8}$ & $2.0^{+0.6}_{-0.6}$ & $5.4^{+1.5}_{-1.5}$\\
Abell 370      & $21.4\pm2.8$ & $1342^{+80}_{-85}$ & $5.8\pm0.8$ & $570^{+36}_{-32}$ & $3.9^{+0.8}_{-0.7}$ & $1453^{+59}_{-68}$ & $12.9^{+2.3}_{-2.4}$ & $28.6^{+6.7}_{-6.0}$ & $5.2^{+1.2}_{-1.1}$ & $15.6^{+3.7}_{-3.3}$\\
Abell 383      & $11.0\pm3.4$ & $769^{+111}_{-128}$ & $2.0\pm0.6$ & $351^{+64}_{-43}$ & $0.7^{+0.5}_{-0.3}$ & $1059^{+133}_{-140}$ & $4.1^{+1.9}_{-1.6}$ & $4.3^{+2.0}_{-2.0} $ & $0.9^{+0.4}_{-0.4}$ & $2.4^{+1.1}_{-1.1}$\\
Abell 963      & $13.6\pm2.6$ & $893^{+78}_{-85}$ & $2.4\pm0.5$ & $420^{+35}_{-39}$ & $1.3^{+0.4}_{-0.4}$ & $1020^{+94}_{-102}$ & $3.7^{+1.4}_{-1.3}$ & $7.9^{+2.1}_{-2.3}$ & $1.6^{+0.4}_{-0.5}$ & $4.4^{+1.2}_{-1.3}$\\
Abell 1689     & $31.1\pm3.0$ & $1298^{+59}_{-61}$ & $6.3\pm0.6$ & $659^{+25}_{-25}$ & $4.9^{+0.6}_{-0.6}$ & $1594^{+87}_{-91}$ & $13.9^{+2.6}_{-2.5}$ & $26.4^{+4.7}_{-4.3}$ & $4.9^{+0.9}_{-0.8}$ & $14.1^{+2.5}_{-2.3}$ \\
Abell 1763     & $19.2\pm2.9$ & $1075^{+77}_{-82}$ & $4.4\pm0.5$ & $554^{+27}_{-29}$ & $3.0^{+0.5}_{-0.5}$ & $1422^{+104}_{-104}$ & $10.2^{+2.7}_{-2.4}$ & $13.6^{+3.3}_{-3.3}$ & $2.7^{+0.6}_{-0.6}$ & $7.5^{+1.8}_{-1.8}$\\
Abell 2218     & $19.5\pm3.1$ & $1024^{+76}_{-82}$ & $3.6\pm0.6$ & $524^{+38}_{-40}$ & $2.4^{+0.6}_{-0.6}$ & $1155^{+81}_{-80}$ & $5.2^{+1.4}_{-1.3}$ & $11.8^{+3.1}_{-2.9}$ & $2.4^{+0.6}_{-0.6}$ & $6.5^{+1.7}_{-1.6}$\\
Abell 2219     & $18.5\pm2.7$ & $1067^{+70}_{-75}$ & $4.2\pm0.6$ & $545^{+32}_{-40}$ & $2.9^{+0.6}_{-0.6}$ & $1374^{+76}_{-75}$ & $9.3^{+2.0}_{-1.9}$ & $14.3^{+3.4}_{-3.0}$ & $2.8^{+0.7}_{-0.6}$ & $7.8^{+1.9}_{-1.6}$\\
Abell 2390     & $23.2\pm2.6$ & $1171^{+61}_{-64}$ & $4.3\pm0.5$ & $555^{+24}_{-26}$ & $3.1^{+0.5}_{-0.5}$ & $1345^{+62}_{-61}$ & $8.7^{+1.7}_{-1.6}$ & $20.7^{+3.7}_{-3.6}$ & $3.9^{+0.7}_{-0.7}$ & $11.2^{+2.0}_{-1.9}$\\
MS 0015.9+1609 & $14.1\pm2.7$ & $1356^{+118}_{-129}$ & $6.6\pm0.8$ & $567^{+34}_{-35}$ & $4.7^{+0.9}_{-0.9}$ & $1626^{+59}_{-66}$ & $22.2^{+3.9}_{-4.0}$ & $26.8^{+8.5}_{-7.5}$ & $4.8^{+1.5}_{-1.3}$ & $14.8^{+4.7}_{-4.1}$\\
MS 0906.5+1110 & $16.0\pm2.8$ & $906^{+73}_{-79}$ & $3.0\pm0.5$ & $477^{+25}_{-32}$ & $1.8^{+0.4}_{-0.4}$ & $1380^{+76}_{-98}$ & $8.9^{+1.9}_{-2.0}$ & $9.3^{+2.4}_{-2.3}$ & $1.9^{+0.5}_{-0.5}$ & $5.1^{+1.3}_{-1.3}$\\
MS 1224.7+2007 & $8.6\pm2.8$ & $817^{+121}_{-141}$ & $2.5\pm0.7$ & $388^{+47}_{-77}$ & $1.2^{+0.5}_{-0.6}$ & $881^{+77}_{-76}$ & $2.7^{+1.1}_{-1.0}$ & $4.7^{+2.5}_{-2.1}$ & $1.0^{+0.5}_{-0.4}$ & $2.7^{+1.4}_{-1.2}$\\
MS 1231.3+1542 & $5.1\pm2.6$ & $549^{+118}_{-149}$ & $0.8\pm0.5$ & $295^{+40}_{-44}$ & $0.5^{+0.2}_{-0.2}$ & $547^{+93}_{-125}$ & $0.6^{+0.5}_{-0.4}$ & $1.4^{+1.3}_{-0.9}$ & $0.3^{+0.3}_{-0.2}$ & $0.8^{+0.8}_{-0.5}$\\
MS 1358.4+6245 & $12.8\pm2.6$ & $994^{+90}_{-99}$ & $3.6\pm0.6$ & $471^{+27}_{-35}$ & $2.1^{+0.4}_{-0.5}$ & $1140^{+88}_{-85}$ & $5.9^{+1.9}_{-1.7}$ & $10.7^{+3.4}_{-3.0}$ & $2.1^{+0.7}_{-0.6}$ & $6.0^{+1.9}_{-1.7}$\\
MS 1455.0+2232 & $14.6\pm2.3$ & $945^{+67}_{-72}$ & $2.7\pm0.5$ & $423^{+32}_{-29}$ & $1.4^{+0.4}_{-0.3}$ & $1051^{+49}_{-61}$ & $4.3^{+1.2}_{-1.2}$ & $8.6^{+1.9}_{-2.0}$ & $1.8^{+0.4}_{-0.4}$ & $4.8^{+1.0}_{-1.1}$\\
MS 1512.4+3647 & $7.7\pm2.7$ & $786^{+120}_{-140}$ & $1.8\pm0.7$ & $326^{+40}_{-54}$ & $0.7^{+0.3}_{-0.3}$ & $859^{+158}_{-215}$ & $2.7^{+1.9}_{-1.7}$ & $4.3^{+2.0}_{-2.0}$ & $0.9^{+0.4}_{-0.4}$ & $2.5^{+1.2}_{-1.2}$\\
MS 1621.5+2640 & $15.0\pm2.7$ & $1186^{+99}_{-108}$ & $4.2\pm0.7$ & $476^{+49}_{-73}$ & $2.4^{+0.8}_{-1.0}$ & $1205^{+66}_{-73}$ & $7.8^{+2.0}_{-2.0}$ & $16.8^{+5.1}_{-4.5}$ & $3.2^{+1.0}_{-0.9}$ & $9.4^{+2.8}_{-2.5}$ \\
CL0024.0+1652  & $16.9\pm3.0$ & $1242^{+101}_{-110}$ & $5.2\pm0.8$ & $545^{+30}_{-38}$ & $3.5^{+0.7}_{-0.7}$ & $1323^{+95}_{-111}$ & $9.9^{+2.8}_{-2.8}$ & $20.4^{+6.1}_{-5.5}$ & $3.8^{+1.1}_{-1.0}$ & $11.2^{+3.3}_{-3.0}$\\
\hline
Abell 115N   & $11.9\pm2.8$ & $814^{+88}_{-98}$ & $1.7\pm0.5$ & $332^{+55}_{-87}$ & $0.6^{+0.4}_{-0.4}$ & $1041^{+99}_{-143}$ & $3.9^{+1.4}_{-1.5}$ & $5.7^{+1.7}_{-1.9}$ & $1.2^{+0.4}_{-0.4}$ & $3.2^{+1.0}_{-1.0}$\\
Abell 115S   & $12.0\pm2.8$ & $819^{+85}_{-95}$ & $2.4\pm0.5$ & $355^{+82}_{-43}$ & $0.8^{+0.7}_{-0.3}$ & $1157^{+71}_{-72}$ & $5.4^{+1.3}_{-1.2}$ & $6.8^{+2.1}_{-1.9}$ & $1.4^{+0.4}_{-0.4}$ & $3.8^{+1.2}_{-1.1}$\\
Abell 222    & $15.1\pm2.9$ & $934^{+82}_{-90}$ & $3.0\pm0.5$ & $451^{+45}_{-54}$ & $1.6^{+0.6}_{-0.5}$ & $1181^{+68}_{-70}$ & $5.8^{+1.3}_{-1.2}$ & $7.9^{+2.4}_{-2.1}$ & $1.6^{+0.5}_{-0.4}$ & $4.4^{+1.4}_{-1.2}$\\
Abell 223N   & $16.7\pm2.9$ & $976^{+78}_{-85}$ & $3.0\pm0.5$ & $441^{+47}_{-87}$ & $1.5^{+0.5}_{-0.7}$ & $1235^{+83}_{-78}$ & $6.6^{+1.6}_{-1.4}$ & $9.7^{+2.8}_{-2.4}$ & $2.0^{+0.6}_{-0.5}$ & $5.3^{+1.5}_{-1.3}$\\
Abell 223S   & $11.3\pm3.0$ & $805^{+96}_{-109}$ & $2.5\pm0.5$ & $388^{+57}_{-56}$ & $1.0^{+0.5}_{-0.4}$ & $1261^{+107}_{-96}$ & $7.0^{+2.1}_{-1.7}$ & $5.7^{+2.0}_{-2.1}$ & $1.2^{+0.4}_{-0.5}$ & $3.2^{+1.1}_{-1.2}$ \\
Abell 520    & $20.4\pm2.7$ & $1064^{+66}_{-70}$ & $3.4\pm0.5$ & $495^{+28}_{-31}$ & $2.1^{+0.4}_{-0.4}$ & $1176^{+72}_{-69}$ & $5.6^{+1.3}_{-1.2}$ & $13.2^{+2.5}_{-2.8}$ & $2.6^{+0.5}_{-0.6}$ & $7.2^{+1.4}_{-1.5}$\\
Abell 521    & $11.8\pm2.9$ & $869^{+98}_{-110}$ & $3.0\pm0.5$ & $420^{+57}_{-88}$ & $1.4^{+0.6}_{-0.7}$ & $1202^{+88}_{-75}$ & $6.4^{+1.7}_{-1.4}$ & $8.6^{+2.6}_{-2.6}$ & $1.8^{+0.5}_{-0.5}$ & $4.8^{+1.4}_{-1.4}$\\
Abell 586    & $14.1\pm3.2$ & $868^{+92}_{-103}$ & $2.5\pm0.6$ & $420^{+48}_{-49}$ & $1.3^{+0.5}_{-0.4}$ & $1192^{+91}_{-100}$ & $5.7^{+1.6}_{-1.5}$ & $6.8^{+2.4}_{-2.2}$ & $1.4^{+0.5}_{-0.5}$ & $3.8^{+1.3}_{-1.2}$\\
Abell 611    & $12.8\pm2.9$ & $954^{+99}_{-110}$ & $3.3\pm0.7$ & $459^{+44}_{-58}$ & $1.8^{+0.6}_{-0.6}$ & $1146^{+65}_{-62}$ & $5.8^{+1.3}_{-1.2}$ & $8.2^{+2.8}_{-2.5}$ & $1.7^{+0.6}_{-0.5}$ & $4.6^{+1.6}_{-1.4}$\\
Abell 697    & $16.6\pm2.7$ & $1059^{+81}_{-87}$ & $4.1\pm0.5$ & $518^{+32}_{-36}$ & $2.6^{+0.5}_{-0.5}$ & $1371^{+49}_{-56}$ & $9.8^{+1.5}_{-1.5}$ & $11.4^{+3.0}_{-2.7}$ & $2.3^{+0.6}_{-0.5}$ & $6.4^{+1.7}_{-1.5}$\\
Abell 851    & $18.1\pm3.0$ & $1272^{+98}_{-105}$ & $5.1\pm0.5$ & $522^{+18}_{-22}$ & $3.1^{+0.4}_{-0.4}$ & $1343^{+85}_{-88}$ & $10.6^{+2.5}_{-2.3}$ & $20.0^{+5.4}_{-5.0}$ & $3.8^{+1.0}_{-0.9}$ & $11.1^{+3.0}_{-2.8}$ \\
Abell 959    & $21.8\pm2.9$ & $1217^{+77}_{-82}$ & $4.7\pm0.6$ & $554^{+29}_{-32}$ & $3.2^{+0.6}_{-0.6}$ & $1276^{+72}_{-71}$ & $7.9^{+1.7}_{-1.6}$ & $18.6^{+4.0}_{-3.9}$ & $3.6^{+0.8}_{-0.7}$ & $10.2^{+2.2}_{-2.1}$ \\
Abell 1234   & $17.3\pm2.8$ & $938^{+69}_{-74}$ & $2.3\pm0.5$ & $423^{+29}_{-31}$ & $1.3^{+0.3}_{-0.3}$ & $980^{+96}_{-94}$ & $3.1^{+1.2}_{-1.0}$ & $7.9^{+1.9}_{-2.0}$ & $1.6^{+0.4}_{-0.4}$ & $4.4^{+1.0}_{-1.1}$ \\
Abell 1246   & $14.4\pm2.6$ & $880^{+71}_{-77}$ & $2.5\pm0.4$ & $409^{+36}_{-43}$ & $1.2^{+0.4}_{-0.4}$ & $1080^{+53}_{-80}$ & $4.3^{+0.9}_{-1.1}$ & $8.2^{+1.8}_{-1.9}$ & $1.7^{+0.4}_{-0.4}$ & $4.6^{+1.0}_{-1.1}$\\
Abell 1758E  & $17.1\pm2.4$ & $1062^{+68}_{-72}$ & $4.2\pm0.5$ & $530^{+27}_{-31}$ & $2.8^{+0.5}_{-0.5}$ & $1391^{+92}_{-78}$ & $10.2^{+2.4}_{-1.9}$ & $12.9^{+2.6}_{-2.6}$ & $2.5^{+0.5}_{-0.5}$ & $7.1^{+1.4}_{-1.4}$\\
Abell 1758W  & $21.0\pm2.4$ & $1175^{+60}_{-63}$ & $4.1\pm0.5$ & $522^{+25}_{-29}$ & $2.7^{+0.4}_{-0.5}$ & $1388^{+52}_{-60}$ & $10.1^{+1.6}_{-1.6}$ & $16.4^{+2.7}_{-2.9}$ & $3.2^{+0.5}_{-0.6}$ & $9.0^{+1.5}_{-1.6}$\\
Abell 1835   & $23.2\pm2.9$ & $1218^{+70}_{-74}$ & $4.6\pm0.5$ & $560^{+25}_{-27}$ & $3.2^{+0.5}_{-0.5}$ & $1322^{+51}_{-52}$ & $8.5^{+1.4}_{-1.4}$ & $18.9^{+3.8}_{-3.5}$ & $3.6^{+0.7}_{-0.7}$ & $10.3^{+2.1}_{-1.9}$\\
Abell 1914   & $21.7\pm2.5$ & $1057^{+55}_{-57}$ & $3.5\pm0.4$ & $509^{+20}_{-21}$ & $2.2^{+0.3}_{-0.3}$ & $1193^{+55}_{-56}$ & $5.7^{+1.1}_{-1.1}$ & $13.2^{+2.2}_{-2.1}$ & $2.6^{+0.4}_{-0.4}$ & $7.2^{+1.2}_{-1.1}$\\
Abell 1942   & $14.4\pm2.6$ & $916^{+75}_{-81}$ & $3.1\pm0.4$ & $462^{+25}_{-25}$ & $1.8^{+0.3}_{-0.3}$ & $1070^{+61}_{-69}$ & $4.4^{+1.1}_{-1.0}$ & $8.9^{+2.1}_{-1.9}$ & $1.8^{+0.4}_{-0.4}$ & $5.0^{+1.2}_{-1.1}$\\
Abell 2104   & $23.0\pm3.6$ & $1067^{+80}_{-86}$ & $3.4\pm0.5$ & $513^{+34}_{-45}$ & $2.2^{+0.5}_{-0.6}$ & $1235^{+84}_{-80}$ & $6.2^{+1.6}_{-1.4}$ & $13.2^{+3.5}_{-3.1}$ & $2.6^{+0.7}_{-0.6}$ & $7.2^{+1.9}_{-1.7}$\\
Abell 2111   & $16.0\pm2.7$ & $974^{+74}_{-79}$ & $3.4\pm0.5$ & $489^{+28}_{-29}$ & $2.1^{+0.4}_{-0.4}$ & $1087^{+109}_{-87}$ & $4.6^{+1.7}_{-1.2}$ & $9.7^{+1.9}_{-2.2}$ & $2.0^{+0.4}_{-0.5}$ & $5.4^{+1.1}_{-1.2}$\\
Abell 2163   & $21.2\pm3.2$ & $1094^{+80}_{-86}$ & $4.2\pm0.6$ & $556^{+38}_{-42}$ & $3.0^{+0.7}_{-0.7}$ & $1403^{+113}_{-102}$ & $9.6^{+2.7}_{-2.2}$ & $14.7^{+3.5}_{-3.4}$ & $2.9^{+0.7}_{-0.7}$ & $8.0^{+1.9}_{-1.8}$\\
Abell 2204   & $25.4\pm3.1$ & $1128^{+65}_{-69}$ & $4.4\pm0.5$ & $583^{+29}_{-31}$ & $3.3^{+0.6}_{-0.6}$ & $1357^{+73}_{-70}$ & $8.3^{+1.6}_{-1.5}$ & $15.7^{+3.0}_{-3.0}$ & $3.1^{+0.6}_{-0.6}$ & $8.5^{+1.6}_{-1.6}$\\
Abell 2259   & $13.3\pm3.0$ & $830^{+87}_{-97}$ & $1.9\pm0.5$ & $348^{+43}_{-38}$ & $0.7^{+0.3}_{-0.2}$ & $1063^{+70}_{-120}$ & $4.0^{+1.1}_{-1.4}$ & $6.1^{+2.2}_{-1.8}$ & $1.3^{+0.5}_{-0.4}$ & $3.4^{+1.2}_{-1.0}$\\
Abell 2261   & $22.2\pm2.8$ & $1153^{+69}_{-73}$ & $5.0\pm0.5$ & $605^{+26}_{-28}$ & $3.9^{+0.6}_{-0.6}$ & $1540^{+54}_{-46}$ & $13.0^{+1.8}_{-1.6}$ & $17.5^{+3.2}_{-3.2}$ & $3.4^{+0.6}_{-0.6}$ & $9.5^{+1.8}_{-1.8}$\\
Abell 2537   & $18.7\pm2.8$ & $1150^{+80}_{-86}$ & $4.9\pm0.5$ & $557^{+24}_{-26}$ & $3.3^{+0.5}_{-0.5}$ & $1235^{+50}_{-50}$ & $7.3^{+1.3}_{-1.3}$ & $16.8^{+3.8}_{-3.5}$ & $3.2^{+0.7}_{-0.7}$ & $9.2^{+2.1}_{-1.9}$\\
MS0440.5+0204    & $11.5\pm3.2$ & $798^{+103}_{-117}$ & $2.5\pm0.6$ & $419^{+46}_{-63}$ & $1.3^{+0.5}_{-0.5}$ & $867^{+56}_{-56}$ & $2.2^{+0.7}_{-0.7}$ & $3.9^{+1.8}_{-1.6}$ & $0.9^{+0.4}_{-0.4}$ & $2.2^{+1.0}_{-0.9}$\\ 
MS0451.6-0305     & $9.5\pm3.1$ & $1086^{+162}_{-188}$ & $3.6\pm0.8$ & $412^{+41}_{-42}$ & $1.8^{+0.6}_{-0.5}$ & $964^{+91}_{-108}$ & $4.6^{+1.6}_{-1.6}$ & $11.1^{+5.2}_{-4.5}$ & $2.2^{+1.0}_{-0.9}$ & $6.3^{+3.0}_{-2.6}$\\
MS1008.1-1224     & $18.4\pm3.1$ & $1155^{+93}_{-101}$ & $3.5\pm0.6$ & $475^{+33}_{-35}$ & $2.1^{+0.5}_{-0.4}$ & $1080^{+41}_{-63}$ & $4.9^{+1.0}_{-1.1}$ & $15.0^{+4.0}_{-3.7}$ & $2.9^{+0.8}_{-0.7}$ & $8.3^{+2.2}_{-2.1}$\\
RXJ1347.5-1145    & $16.1\pm3.0$ & $1271^{+111}_{-121}$ & $4.8\pm0.7$ & $502^{+38}_{-40}$ & $2.9^{+0.7}_{-0.7}$ & $1270^{+100}_{-144}$ & $9.5^{+2.7}_{-3.1}$ & $18.2^{+5.9}_{-5.1}$ & $3.4^{+1.1}_{-1.0}$ & $10.1^{+3.3}_{-2.8}$\\
RXJ1524.6+0957    & $7.3\pm2.8$ & $914^{+157}_{-189}$ & $2.8\pm0.9$ & $344^{+59}_{-67}$ & $1.0^{+0.6}_{-0.5}$ & $883^{+112}_{-129}$ & $3.4^{+1.6}_{-1.4}$ & $7.5^{+3.6}_{-3.4}$ & $1.5^{+0.7}_{-0.7}$ & $4.3^{+2.1}_{-1.9}$\\
MACS J0717.5+3745 & $21.1\pm3.2$ & $1648^{+117}_{-125}$ & $7.2\pm1.0$ & $612^{+40}_{-51}$ & $5.9^{+1.3}_{-1.4}$ & $1483^{+74}_{-76}$ & $16.9^{+3.2}_{-3.0}$ & $51.4^{+11.3}_{-10.8}$ & $8.7^{+1.9}_{-1.8}$ & $27.9^{+6.1}_{-5.9}$\\
MACS J0913.7+4056 & $9.8\pm3.2$ & $974^{+145}_{-169}$ & $3.5\pm0.8$ & $413^{+63}_{-71}$ & $1.6^{+0.9}_{-0.7}$ & $962^{+71}_{-81}$ & $4.1^{+1.2}_{-1.2}$ & $7.9^{+4.0}_{-3.4}$ & $1.6^{+0.8}_{-0.7}$ & $4.5^{+2.3}_{-2.0}$\\
CIZA J1938+54     & $16.3\pm2.9$ & $1037^{+86}_{-93}$ & $4.3\pm0.5$ & $529^{+22}_{-23}$ & $2.8^{+0.4}_{-0.4}$ & $1350^{+89}_{-96}$ & $9.1^{+2.2}_{-2.1}$ & $11.4^{+3.1}_{-3.0}$ & $2.3^{+0.6}_{-0.6}$ & $6.3^{+1.7}_{-1.7}$\\
3C295             & $11.8\pm2.6$ & $1085^{+107}_{-118}$ & $4.5\pm0.7$ & $486^{+33}_{-37}$ & $2.7^{+0.6}_{-0.6}$ & $1075^{+52}_{-62}$ & $5.8^{+1.2}_{-1.3}$ & $13.2^{+4.1}_{-3.6}$ & $2.6^{+0.8}_{-0.7}$ & $7.4^{+2.3}_{-2.0}$\\
\hline
\hline
\end{tabular}
\bigskip
\begin{minipage}{\linewidth}
{\footnotesize Column 2: cluster name; Column 3: best fit
Einstein radius for the SIS model; Column~4: line-of-sight
velocity dispersion of the best fit SIS model; Column~5: 
projected mass within an aperture of $0.5h_{70}^{-1}$Mpc;
Columns~7 \& 9: $r_{\Delta}$ determined using aperture masses;
Columns~8 \& 10: deprojected aperture masses within $r_\Delta$;
Columns~11-13: masses from best fit NFW model. All masses
are listed in units of $[10^{14}h_{70}^{-1}$\msun]}
\end{minipage}
\end{center}
\end{table*}

\section{Mass determination}

One of the advantages of weak gravitational lensing is that it
provides a direct measure of the projected mass distribution, without
having to make assumptions about the density profile. In practice,
however, the conversion of the observations into useful mass estimates
does depend on the density profile, although the dependence can be
weak.  Furthermore, projected masses cannot be compared directly to
results from other techniques, such as X-ray
observations. Alternatively one can assume a parametric form for the
density distribution and fit the predicted tangential distortion to
the data. We discuss and compare the results from both approaches in
the following sections.

\subsection{Parametric mass model}

Numerical simulations of cold dark matter indicate that the density
profiles of dark matter halos are well described by the fitting
function proposed by \cite{NFW96, NFW}. The profile is characterized by two
parameters: the mass of the halo and the concentration $c$ (or
characteristic scale). A commonly used method to infer cluster masses
is to compare this model to the observed lensing signal and to
determine the best fit values for the mass and the concentration,
although $c$ is typically poorly constrained. In simulations, however,
the mass and the concentration are found to be correlated, albeit with
an intrinsic scatter, which reflects the variation in halo formation
histories. We follow the definitions of \cite{Hoekstra07}, but use a
more recent relation between the virial mass $M_{\rm vir}$ and
concentration $c$, which was determined by \cite{Duffy08} based on the
cosmological parameters that best fit the WMAP5 observations
\citep{Komatsu09}:

\begin{equation}
c=7.85\left(\frac{M^{\rm NFW}_{\rm vir}}{2\times 10^{12}}\right)^{-0.081} (1+z)^{-0.71}.
\label{cmrel}
\end{equation}

\noindent This relation is used when we fit the NFW model to the
lensing signal at radii $0.5-2h_{70}^{-1}$Mpc. The resulting values
for $M^{\rm NFW}_{\rm vir}$ are presented in Table~\ref{tabmass}.  For
reference, we also list the values for $M^{\rm NFW}_{2500}$ and
$M^{\rm NFW}_{500}$, which are the masses within the radii where the
mean mass density of the halo is respectively 2500 and 500 times the
critical density at the redshift of the cluster, computed from the
best fit virial mass (and Eqn.~\ref{cmrel}). We also fit a SIS model
to the shear at these radii. The resulting Einstein radii and
corresponding velocity dispersions are also listed in
Table~\ref{tabmass}.

\begin{figure}
\begin{center}
\leavevmode
\hbox{%
\epsfxsize=8.5cm
\epsffile{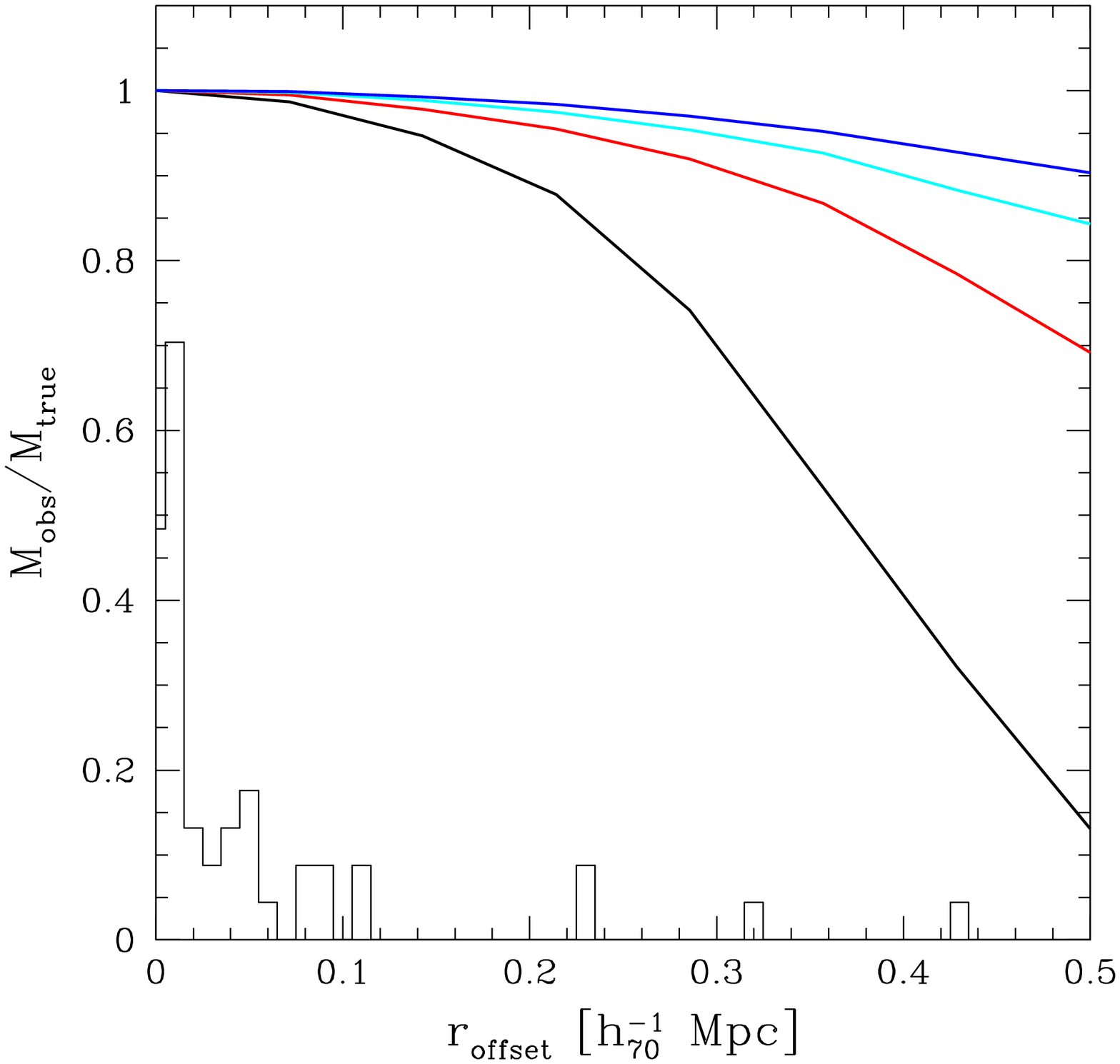}}
\caption{Plot of the ratio of the inferred lensing mass and the true
  mass as a function of centroid offset for an NFW halo with a virial
  mass $M=5\times 10^{14}$\msun. The lensing mass is inferred from the
  $\zeta_c$ statistic as explained in the text. The curves correspond
  to $M_{2500}$ (lowest curve; black), $M_{1000}$ (red), $M_{500}$
  (cyan) and $M_{200}$ (highest curve; blue). For overdensities
  $\Delta<1000$, the observed value for $M_\Delta$ is not very
  sensitive to centroiding errors because the large scale shear is not
  affected.  The histogram shows the distribution of offsets
  between the BCG and the peak of the X-ray emission, suggesting that
  miscentering is a small effect for the CCCP sample.
\label{bias_offset}}
\end{center}
\end{figure}

For the model fit, we avoid both the small and large radii. As
discussed in \cite{Hoekstra01,Hoekstra03,Hoekstra11,Becker11}
structures along the line-of-sight contribute noise to the mass
measurement. This 'cosmic noise' dominates at large radii, where the
cluster signal is small.  We account for the noise introduced by
large-scale structure in the error budget. We do not fit the small
scales in order to reduce (residual) contamination by cluster
members and the effects of substructure (i.e., deviations from the
simple NFW profile). Generally, the presence of substructure in the
cluster core causes a reduction in the tangential shear, which biases
the masses low \citep{Hoekstra02}. As discussed in \S\ref{sec:center},
the choice of cluster center is also important, as an offset from the
true centre biases the mass low as well (also see Fig.~\ref{bias_offset}). The
bias is smaller if the offsets are small relative to the scales on
which the shear is measured \citep[see Figure 4 in][]{Hoekstra11b}.

\subsection{Deprojected aperture mass}\label{sec:apmass}

Rather than adopting a parameterized model that is fit to the data, a
unique feature of weak lensing is that the observed shear can be
related directly to a density contrast. In essence, it allows us to
infer the projected mass within an aperture, with relatively few
assumptions about the actual mass distribution. We use that (Clowe et
al. 1998):

\begin{equation}
\zeta_c(r_1)=2\int_{r_1}^{r_2}d\ln r\langle\gamma_t\rangle+
\frac{2r_{\rm max}^2}{r_{\rm max}^2-r_2^2} \int_{r_2}^{r_{\rm max}}
d\ln r \langle\gamma_t\rangle,
\end{equation}

\noindent which can be expressed in terms of the mean dimensionless
surface density interior to $r_1$ relative to the mean
surface density in an annulus from $r_2$ to $r_{\rm max}$

\begin{equation}
\zeta_c(r_1)=\bar\kappa(r'<r_1)-\bar\kappa(r_2<r'<r_{\rm max}).
\end{equation}

This shows that we can determine the average surface density within a
given aperture up to a constant (i.e., the mean convergence in the
annulus). The surface density in the annulus cannot be ignored, even
though it is small for the wide field imaging data used here. We
estimate the mean surface density in the annulus based on the best fit
NFW model. For the clusters observed with the CFH12k camera, we adopt
$r_2=600''$ and $r_{\rm max}=1000''$, and for the Megacam observations
we use $r_2=900''$ and $r_{\rm max}=1500''$ (reflecting the larger
field-of-view).

Thanks to our ability to measure the lensing signal out to large
radii, the (model dependent) correction to the mass is only $\sim 8\%$
for $M_{500}$ and $\sim 3\%$ for $M_{2500}$. The correction is larger
if we use the best fit SIS model instead of the best NFW fit.  The
difference is largest for $M_{500}$, which would increase by $\sim
10\%$ compared to the results listed in Table~\ref{tabmass}, whereas
$M_{2500}$ would increase by $\sim 4\%$ if we use the SIS fit to
estimate the convergence in the annulus. Note that \cite{Okabe10}
assume that the contribution in the annulus can be ignored, which is
definitely not the case, considering the bias it introduces for a
large sample of clusters. Consequently we cannot compare our results
directly to the aperure masses listed in \cite{Okabe10}.

Although the NFW model should be a good description of the average
mass distribution on small scales, it is less accurate on large scales
because structures near the cluster will contribute to the lensing
signal. This can be studied in the context of a halo-model
\citep[e.g.][]{Johnston07}. However, for the radii and masses we
study here this so-called two-halo contribution is expected to be
small \citep{Johnston07,Becker11}.

We examined the effect of errors in the adopted cluster center on the
inferred aperture masses. The results are presented in
Figure~\ref{bias_offset} for four overdensities $\Delta$. For higher
overdensities one needs to integrate the shear signal to smaller radii
and the mass becomes more sensitive to centroid offsets.  However, the
observed offsets between the BCG position and the peak of the X-ray
emission suggest that the bias in our masses are negligible (even for
$M_{2500}$). Note that for three of the four clusters with offsets
larger than $200h_{70}^{-1}$kpc we use the X-ray center instead (see
\S\ref{sec:center})

To compare to results from other methods the aperture masses need to
be deprojected, under the assumption of spherical
symmetry. Non-parametric deprojections are noisy, and we therefore
employ a different approach, which was also used in \cite{Hoekstra07}.
For each aperture we determine which NFW model would yield the
observed projected mass. The corresponding mass of the model is then
taken to be the deprojected mass. Hence we assume that the density
profile along the line-of-sight is described by the NFW model. Note
that each aperture is treated independently: the corresponding virial
mass (and consequently the concentration given by Eqn.~\ref{cmrel}),
is allowed to vary with radius. The resulting profiles are used to
measure $r_\Delta$. The results for $\Delta=2500$ and $500$ are listed
in Table~\ref{tabmass}, along with the corresponding values for
$r_\Delta$.

\cite{Becker11} have shown that fitting an NFW model to the observed
lensing signal can lead to biased mass estimates when the mass and
concentration are free parameters \citep[also see][]{Meneghetti10,
  Bahe12, High12}. \cite{Becker11} also make the point that the bias
is likely to depend on the method that was used to infer the mass, as
well as the range in angular scales that is considered. In particular,
they show that extending the NFW fits to large radii ($>10'$) tends to
bias the masses low by $\sim 6\%$. Most of the bias appears to be
caused by the fact that the NFW fit overestimates the reduced
tangential shear on scales of $10'-20'$ before two-halo contributions
\citep[see e.g.][]{Johnston07} become important. The NFW fit results
presented in Table~\ref{tabmass} are based on the lensing signal at
radii $0.5-2h_{70}^{-1}$Mpc, for which the bias should be small.

\begin{figure*}
\begin{center}
\leavevmode
\hbox{%
\epsfxsize=8.5cm
\epsffile{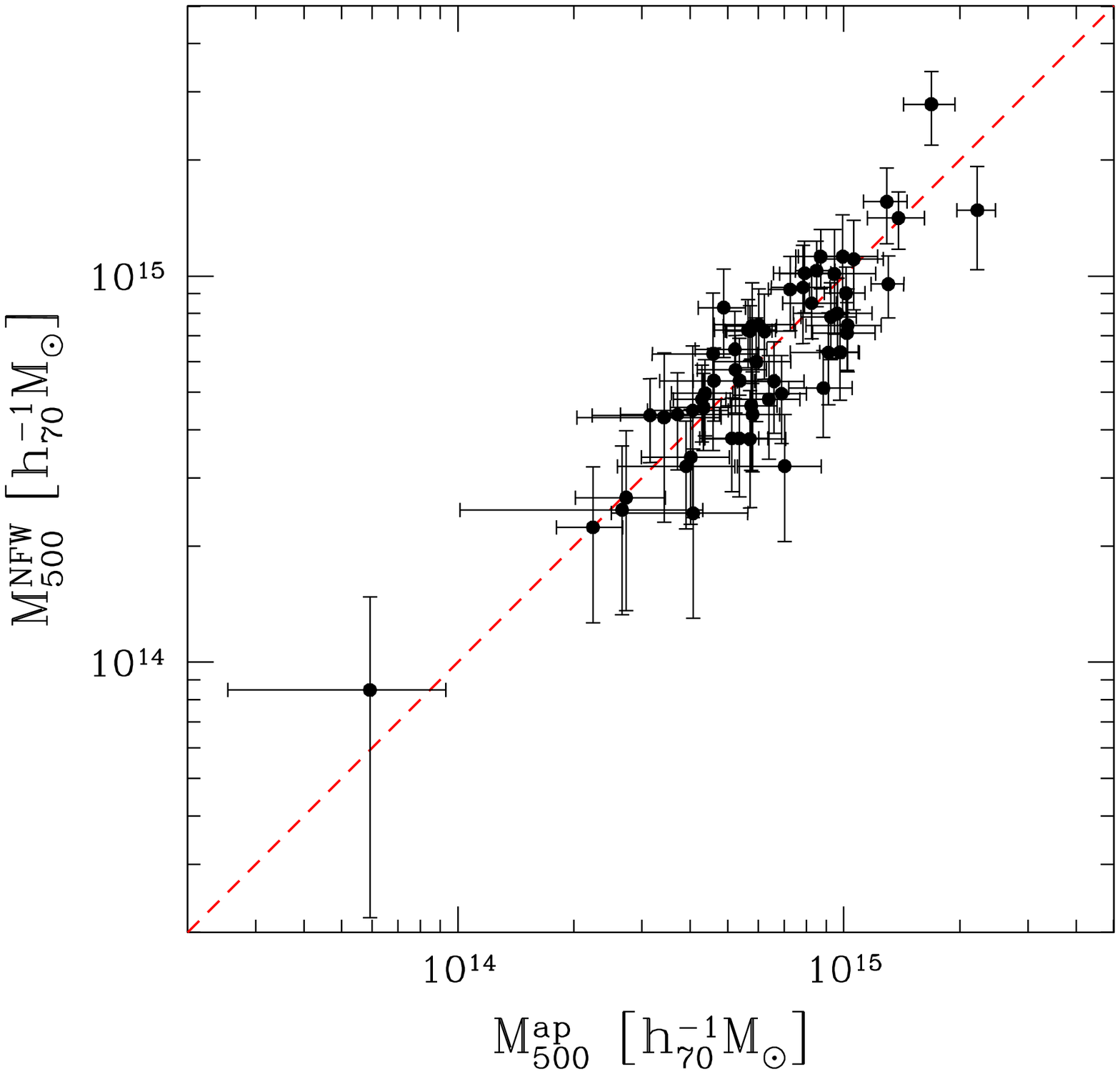}
\epsfxsize=8.5cm
\epsffile{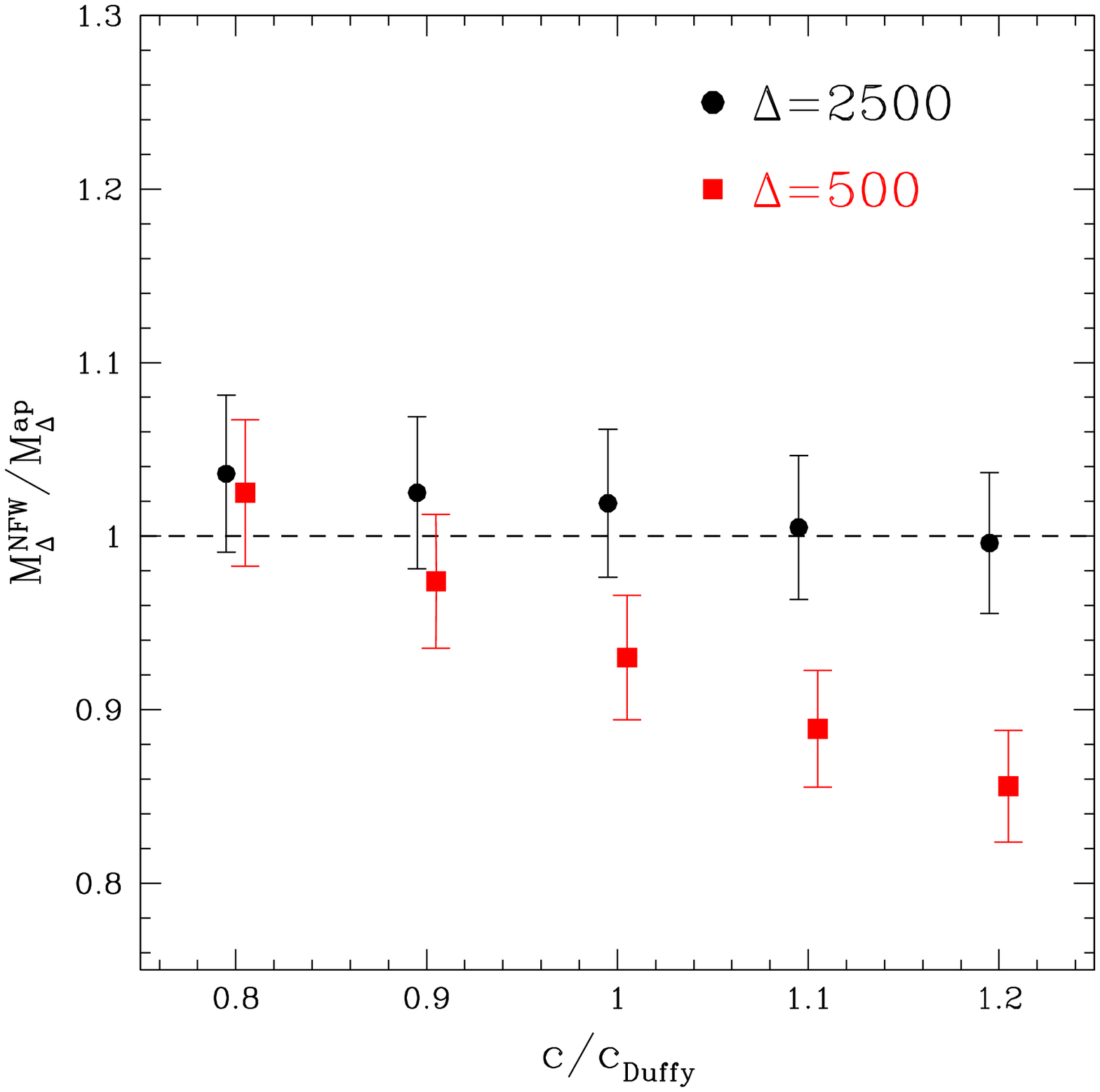}}
\caption{{\it left panel:} Value for $M_{500}$ from the best fit NFW
  model to the lensing signal at radii $0.5-2 h_{70}^{-1}$Mpc versus
  the $M_{500}$ inferred from the deprojected aperture mass. The
  mass-concentration relation from Duffy et al. (2008) was used for
  the NFW fit. The dashed red line indicates the line of equality. {\it
    right panel:} The average ratio of the best fit NFW mass and the
  deprojected aperture mass when the normalization of the
  mass-concentration (Equation~\ref{cmrel}) is changed. High normalisations
  lead to inconsistent estimates for $M_{500}$.
\label{nfw_compare}}
\end{center}
\end{figure*}

The situation for the aperture masses is less clear. \cite{High12}
used simulations to find that their approach underestimates the masses
at $R_{500}$ by $6-13\%$. However, they use the $\zeta$-statistic
proposed by \cite{Fahlman94}, which requires an estimate of $\kappa$
in an annulus ranging from $R_{500}$ out to $16'$, which is the
largest scale considered by \cite{High12}. Compared to our approach,
the correction for the convergence in this annulus is much larger,
about $35\%$ at $R_{500}$, and thus more sensitive to deviations from
the assumed density profile. We expect the bias to be proportional to
this correction, and consequently we expect any bias in our
deprojected masses to be considerably smaller than those of
\cite{High12}.

\begin{table}
\caption{Senstivity to $c(M)$ normalization \label{c_sens}}
\begin{tabular}{ccccc}
\hline
\hline
         & \multicolumn{2}{c}{aperture mass} & \multicolumn{2}{c}{NFW fit}\\
$\Delta$ & $\alpha=0.8$ & $\alpha=1.2$ & $\alpha=0.8$ & $\alpha=1.2$\\
\hline
2500     & $0.88\pm0.03$   & $1.09\pm0.03$    & $0.90\pm0.03$ & $1.07\pm0.03$  \\
1000     & $0.92\pm0.01$   & $1.06\pm0.01$    & $1.00\pm0.04$ & $0.99\pm0.03$  \\
 500     & $0.94\pm0.01$   & $1.04\pm0.01$    & $1.06\pm0.04$ & $0.95\pm0.03$  \\
 vir     & $-$               &  $-$               & $1.16\pm0.05$ & $0.89\pm0.03$  \\
\hline
\hline
\end{tabular}
\bigskip
\begin{minipage}{\linewidth}
{\footnotesize Table of the ratio $\langle M_{\Delta}(\alpha c_{\rm
    Duffy})/M_{\Delta}(c_{\rm Duffy})\rangle$ when the masses are
  estimated from deprojecting the aperture masses (Columns 2 \& 3) or
  by fitting an NFW model to the data (Columns 4 \& 5). The errors correspond to
the standard deviation of the ratios.}
\end{minipage}
\end{table}

\subsection{Sensitivity to mass-concentration relation}

Both procedures we use to derive cluster masses depend on the adopted
mass-concentration relation (Equation~\ref{cmrel}). It is important to
note that $c(M)$ depends on the cosmology: the concentration is
related to the mean density of the Universe when the cluster was
formed \citep[e.g.][]{NFW96, Bullock01}. As more massive halos form
later in a $\Lambda$CDM cosmology, they will have smaller
concentrations \citep[but see][]{Prada11}. Furthermore, the mass
dependence is rather modest when considering a relatively small mass
range, as is done here.

The amplitude, however, is sensitive to the matter density $\Omega_m$
and the normalization of the matter power spectrum $\sigma_8$ and
various fitting functions have been published
\citep[e.g.][]{Bullock01,Neto07,Maccio08,Prada11} in addition to
Equation~\ref{cmrel}. The observed mass-concentration relation can
therefore be used in principle to constrain cosmological parameters
\citep{Ettori10}. Note that the simulations studied by \cite{Prada11}
prefer a significantly higher normalization. This claim appears to be
supported by a number of observational results
\citep[e.g.][]{Zitrin10,Zitrin11,Umetsu11}. It is, however, not clear
whether these findings are biased by the fact that many of these are
well-known strong lensing clusters. For instance, accounting for the
3D structure of the clusters appears to yield lower concentrations
\citep{Morandi12}. An important complication, which requires further
study, is the role of baryons, which have not been considered in these
studies. The effect depends on the feedback model, with steeper
profiles if radiative cooling is efficient, and shallower profiles if
feedback is effective \citep[e.g.][]{Lewis00,Duffy10}.

A detailed study of the mass-concentration relation is beyond the
scope of this paper, but it is nonetheless useful to examine the
sensitivity of our mass estimates to our choice for $c(M)$ by varying
its normalization. We studied the average change in the deprojected
aperture masses and the NFW fits and list the results in
Table~\ref{c_sens} for different overdensities. The deprojected mass
converges to the projected mass for large radii and hence it is not
surprising that the aperture mass values for $M_{500}$ are the least
sensitive, with a 20\% variation only changing the masses by $\sim
5\%$. The NFW fits show a similar variation for $M_{500}$ but the
dependence with $c$ is opposite. Interestingly, the value for $M^{\rm
  NFW}_{1000}$ is nearly independent of the adopted normalization,
suggesting that this provides the most robust mass estimate when
fitting an NFW profile to the tangential shear profile between 0.5 and
$2h_{70}^{-1}$Mpc.

If the density profiles are (on average) well described by the NFW
profile we expect good agreement between our different approaches to
measure the cluster masses. The left panel of Figure~\ref{nfw_compare}
shows the results when $M^{\rm NFW}_{500}$ from the NFW fit is compared
to the deprojected aperture mass estimates $M^{\rm ap}_{500}$. Note
that the results are slightly correlated because the NFW fit uses data
out to $2h_{70}^{-1}$Mpc, whereas the typical values for $r_{500}$ are
$\sim 1.2h_{70}^{-1}$Mpc. Hence there is some overlap in the shear
signals that were used to derive the masses. We fit a linear relation
to the masses and find that the ratio of NFW to deprojected aperture
mass is $0.93\pm0.04$.

The right panel of Figure~\ref{nfw_compare} shows how this ratio
varies with the normalization of the mass-concentration relation for
both $M_{2500}$ (black points) and $M_{500}$ (red squares). We find
that for $\Delta=2500$ the agreement is good for a wide range of
normalizations. The comparison of $M_{500}$ suggests a preference for
a lower concentration, but a more detailed analysis is needed before a
conclusion can be drawn. For instance, the presence of substructure
biases the NFW masses low. However, significantly larger
concentrations, such as found by \cite{Prada11} lead to inconsistent
mass estimates, and thus are at odds with our results.

\begin{table*}
\begin{centering}
\caption{SZ measurements\label{tabsz}}
\begin{tabular}{lcccccc}
\hline
\hline
  (1)   &  (2)   & (3) & (4) & (5) & (6) & (7) \\
  name  & $r_{2500}^{\rm SZ}$ & $Y$         & $M^{\rm proj}(<r_{2500}^{\rm SZ})$ 
        & $\Theta^{\rm X}_{500}$ & $Y_{\rm PSX}$ & $M^{\rm proj}(<\Theta_{500}^{\rm X})$\\
        & [$h_{70}^{-1}$kpc]           & $[10^{-10}]$     & $[10^{14}h_{70}^{-1}\msun]$ 
        & [$h_{70}^{-1}$Mpc]           & $[10^{-10}]$     & $[10^{14}h_{70}^{-1}\msun]$\\
\hline
A68     & 616 &	$1.01\pm 0.16$ & $5.0\pm0.8$ & \\
A115N   &     &                &             & 1.22 & $4.27\pm0.58$ & $6.5\pm1.5$\\
A209    &     &                &             & 1.14 & $4.5\pm0.4$ & $8.9\pm1.5$  \\
A267    & 484 &	$0.72\pm 0.10$ & $3.4\pm0.5$ \\
A370    & 508 &	$0.71\pm 0.09$ & $5.9\pm0.8$ \\ 
A520    &     &                &             & 1.24 & $3.9\pm0.5$ & $8.2\pm1.7$ \\
A586    & 529 &	$1.03\pm 0.14$ & $2.7\pm0.6$ \\
A611    & 482 &	$0.54\pm 0.06$ & $3.2\pm0.6$ \\ 
A697    & 568 &	$1.67\pm 0.19$ & $4.8\pm0.5$ & 1.28 & $4.3\pm0.4$ & $12.7\pm1.7$  \\
A963    &     &                &             & 1.11 & $1.6\pm0.3$ & $5.5\pm1.5$\\
A1689   & 664 &	$3.79\pm 0.32$ & $8.1\pm0.8$ & 1.38 & $6.0\pm0.7$ & $16.3\pm1.9$  \\
A1758W  &     &                &             & 1.26 & $2.6\pm0.3$ & $12.8\pm1.8$ \\
A1763   &     &                &             & 1.22 & $3.8\pm0.4$ & $11.2\pm1.7$ \\
A1835   & 672 &	$2.09\pm 0.17$ & $6.5\pm0.8$ \\
A1914   & 660 &	$3.01\pm 0.25$ & $4.6\pm0.7$ & 1.30 & $4.8\pm0.5$ & $8.3\pm1.5$ \\
A2111   & 518 &	$0.95\pm 0.21$ & $3.6\pm0.5$ \\
A2163   & 682 &	$6.89\pm 0.65$ & $6.8\pm0.9$ & 1.52 & $14.6\pm0.6$ & $15.0\pm2.4$ \\ 
A2204   & 671 &	$4.43\pm 0.51$ & $6.2\pm0.9$ & 1.44 & $6.4\pm0.6$  & $12.0\pm1.9$ \\ 
A2218   & 581 &	$1.94\pm 0.19$ & $4.4\pm0.7$ & 1.12 & $3.7\pm0.3$ & $7.1\pm1.4$ \\
A2219   &     &                &             & 1.38 & $7.2\pm0.4$ & $12.8\pm2.0$ \\
A2259   & 476 &	$0.82\pm 0.30$ & $1.8\pm0.5$ \\ 
A2261   & 525 &	$1.34\pm 0.16$ & $5.3\pm0.5$ & 1.33 & $4.1\pm0.4$ & $16.2\pm1.6$ \\ 
A2390   &     &                &             & 1.39 & $4.7\pm0.4$ & $12.3\pm1.9$ \\
MS0015.9+1609 & 507 &	$0.73\pm 0.06$ & $6.7\pm0.8$ \\ 
MS0451.6-0305 & 526 &	$0.66\pm 0.05$ & $3.7\pm0.9$ \\
MS0906.5+1110 &     &                &             & 1.12 & $2.3\pm0.4$ & $9.3\pm1.6$\\ 
MS1358.4+6245 & 539 &	$0.56\pm 0.08$ & $3.9\pm0.6$ \\ 
RX J1347.5-1145   & 706 & $1.61\pm 0.18$ & $7.3\pm1.2$ \\ 
MACS J0717.5+3745 &     &                &             & 1.36 & $2.3\pm0.3$ & $23.5\pm3.7$ \\
CIZA J1938+54     &     &                &             & 1.19 & $2.6\pm0.2$ & $10.8\pm1.8$ \\ 
\hline
\hline
\end{tabular}
\bigskip
\begin{minipage}{\linewidth}
{\footnotesize Column 2: $r_{2500}$ as determined by
  \cite{Bonamente08} from a joint analysis of X-ray and SZ data; Column 3:
  integrated Compton $y-$parameter from \cite{Bonamente08}; Column 4: the
  projected weak lensing mass within $r^{\rm SZ}_{2500}$; Column 5: the
  value for $\Theta_{500}$ from \cite{Planck1}; Column 6: the integrated
  Compton $Y-$parameter within $5\Theta_{500}^{\rm X}$ from \cite{Planck1};
  Column~7: projected weak lensing mass within $\Theta_{500}^{\rm X}$.}
\end{minipage}
\end{centering}
\end{table*}

\section{SZE scaling relation}

On sufficiently large scales, clusters can be considered
representative reservoirs of baryons and dark matter. We therefore
expect correlations between the observable properties and the
underlying mass: more massive systems should have more of
everything. These so-called scaling relations are the result of the
physical processes that give rise to the formation and evolution of
galaxy clusters. If gravity is the dominant process, the scaling
relations and their evolution can be predicted \citep{Kaiser86}.
These self-similar models predict simple power law scaling relations
between the baryonic tracers and the cluster mass. 

Additional non-gravitational processes, such as AGN feedback can in
principle lead to significant deviations from the simple single power
law model \citep[e.g.,][]{Babul02,McCarthy04}. Gravity is, however,
expected to be the dominant process for the masses probed by
CCCP. Furthermore the mass range we study here is limited and we
therefore assume that the scaling relation between the lensing mass
$M_{\rm WL}$ and mass proxy $M_{\rm proxy}$, in our case the
Sunyaev-Zel'dovitch effect (SZE) signal, can be described by a single
power law:

\begin{equation}
M_{\rm WL}=E(z)^\gamma M_0 \left(\frac{M_{\rm proxy}}{M_{\rm pivot}}\right)^\alpha,
\label{eq:powlaw}
\end{equation}

\noindent where $M_0$ is the normalisation, $M_{\rm pivot}$ is the
pivot point, $\alpha$ the power law slope and $E(z)=H(z)/H_0$ is the
normalized Hubble parameter. If we fit Eqn.~\ref{eq:powlaw} to the
measurements in \S5.1 the resulting $\chi^2$ values are larger than
expected, suggesting the presence of intrinsic scatter \citep[also
  see][for a comparison to the X-ray properties]{Mahdavi12}. It is
important to account for the intrinsic scatter when fitting a scaling
relation to the data, because ignoring the scatter will generally bias
the best fit parameters. The presence of intrinsic scatter is also
relevant when assessing the performance of mass-proxies: an observable
with larger statistical errors but smaller intrinsic scatter may be
preferable over one with small statistical errors but a large
intrinsic scatter.

Some of this scatter arises from the fact that the lensing signal
measures the projected mass along the line-of-sight, whereas for
instance the X-ray luminosity probes the virialized regions of the
cluster. Hence, the lensing mass is more sensitive to the fact that
dark matter halos are triaxial. The separation of this `geometric'
source of scatter from the physical scatter requires a careful
comparison with numerical simulations, incorporating in detail the
steps taken in the analysis of the observations.  The effects of
triaxiality on weak lensing mass measurements have been studied using
numerical simulations \citep[e.g.,][]{Metzler01,Meneghetti10,Becker11}
and analytically \citep{Corless07}. The studies by \cite{Corless07} and
\cite{Meneghetti10} suggest a contribution of $\sigma_M\sim 0.15
M_{\rm WL}$ due to the triaxiality of the cluster dark matter halo.

\subsection{Fitting procedure}

We fit scaling relations to the actual measurements (not the
logarithm), because they follow a (close to) normal distribution, even
if the relative error is large. The intrinsic scatter, however, is
assumed to be log-normal, but relatively small. It can therefore be
approximated by a normal distribution with a dispersion
$\sigma_{Q}\approx \ln(10) Q \sigma_{\log Q}$ (we use the log with
base 10).

A number of techniques exist to fit a model to data with errors in
both directions, but we follow a maximum likelihood approach (also see
\cite{Weiner06} for a detailed discussion). For a model $f$ with
parameters ${\bf a}$, the predicted values are $y_i=f(x_i;{\bf
  a})$. The uncertainties in $x_i$ and $y_i$ are given by
$\sigma_{x,i}$ and $\sigma_{y,i}$. If we assume a Gaussian intrinsic
scatter $\sigma_Q$ in the $y$ coordinate, the likelihood ${\cal L}$ is
given by

\begin{equation}
{\cal L}=\prod_{i=1}^{n}\frac{1}{\sqrt{2\pi}w_i}\exp\left[-\frac{[y_i-f(x_i;{\bf a})]^2}{2w_i^2}\right],
\end{equation}

\noindent where $w_i$ accounts for the scatter:

\begin{equation}
w_i^2=\left[{\frac{df}{dx}(x_i)}\right]^2\sigma_{x,i}^2+\sigma_{y,i}^2+\sigma_Q^2.
\end{equation}

\noindent If we consider the logarithm of the likelihood it becomes clear
why including the intrinsic scatter differs from standard least squares
minimization:

\begin{equation}
-2\ln {\cal L}=2\sum_{i=1}^n \ln w_i + \sum_{i=1}^n \left(\frac{y_i-f(x_i;a_j)}{w_i}\right)^2+C,
\end{equation}

\noindent where the second term corresponds to the usual $\chi^2$ and
$C$ is a constant. If there is no intrinsic scatter the first term is
a constant for a given data set and the likelihood is maximized by
minimizing $\chi^2$. However, the first term cannot be ignored if
intrinsic scatter is included as a free parameter.

\begin{figure*}
\begin{center}
\leavevmode \hbox{%
\epsfxsize=8.5cm \epsffile{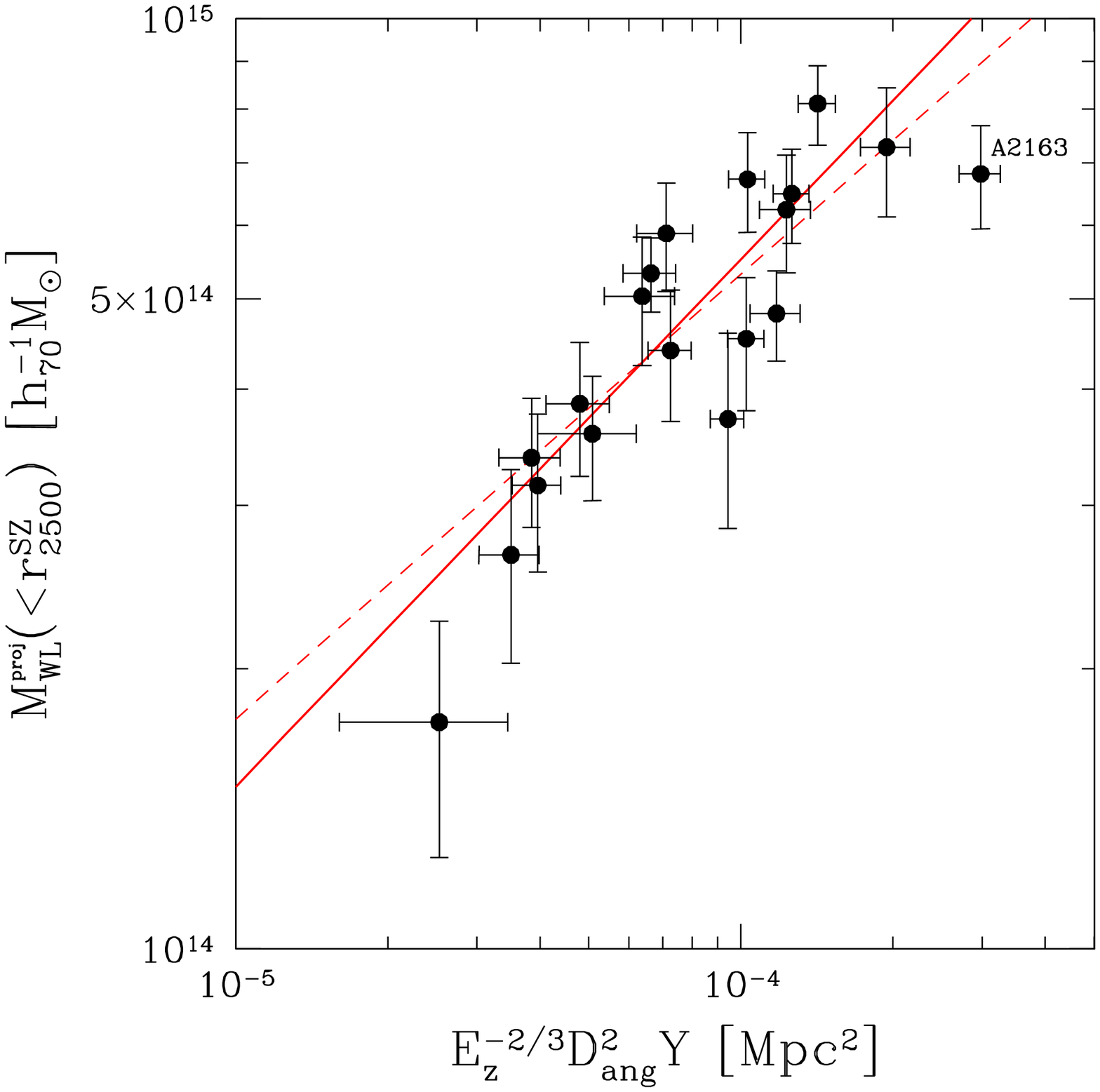}
\epsfxsize=8.5cm \epsffile{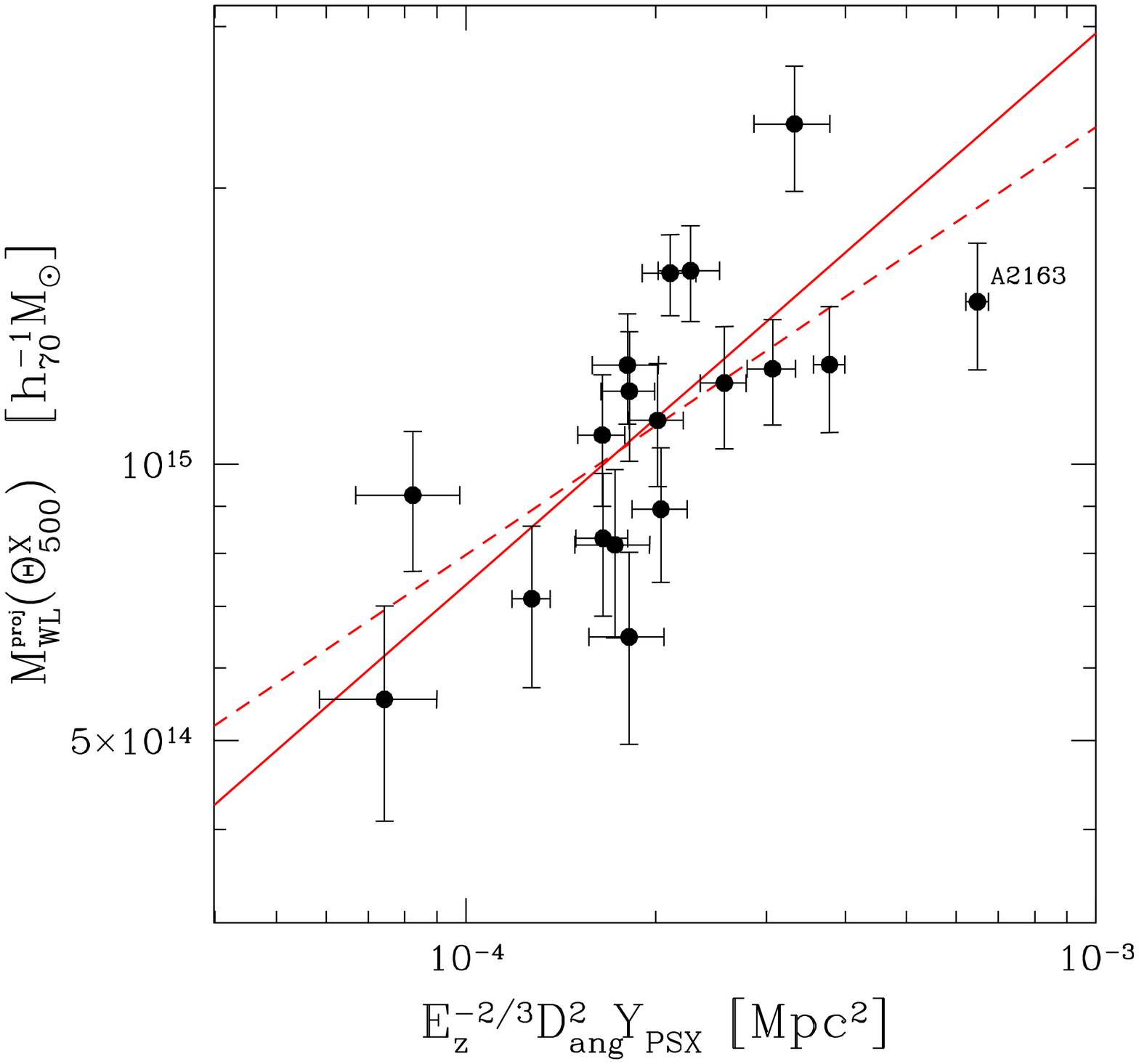}}
\caption{{\it left panel:} Plot of the projected weak lensing mass as
  a function of SZ signal. The mass is measured within an aperture of
  radius $r^{\rm SZ}_{2500}$ as determined by Bonamente et al. (2008)
  based on a joint analysis of X-ray and SZ data. The solid red line
  indicates the best fit power law model when the merging cluster
  A2163 (indicated) is excluded, whereas the dashed line is for the
  full sample. {\it Right panel:} Weak lensing mass within
  $\Theta_{500}^X$ as a function of projected $Y$ from Planck
  Collaboration et al. (2011).
  \label{m_sz}}
\end{center}
\end{figure*}

\subsection{SZE measurements}

The inverse Compton scattering of Cosmic Microwave Background (CMB)
photons by the hot electrons in the intracluster medium results in a
small distortion of the CMB spectrum, known as the Sunyaev-Zel'dovich
effect \citep[SZE;][]{SZ70}. Depending on the observed frequency this
leads to a reduction or increase in the brightness of the CMB at the
location of the cluster \citep[see e.g.][]{Birkinshaw99}. The surface
brightness of the effect does not depend on redshift. As a
consequence, it is (almost) as easy to detect high redshift clusters
as it is to find them at lower redshifts. It is for this reason that
the SZE has developed into an important technique to search for clusters
of galaxies with the aim to constrain cosmological parameters
\citep[e.g.][]{Carlstrom02}. A number of dedicated surveys have
started to release results \citep[e.g.,][]{Williamson11,Marriage11}.

An additional benefit of SZ observations is the fact that the
amplitude of the SZE, quantified by the integrated Compton
$y-$parameter $Y$, is expected to be a good measure of the cluster
mass. It is, however, still sensitive to the cluster physics
\citep[e.g.,][]{McCarthy03a,McCarthy03b} and thus needs to be calibrated
observationally, ideally using weak gravitational lensing. For
instance \cite{High12} present the first weak lensing masses from
dedicated follow-up observations for clusters discovered by the South
Pole Telescope (SPT; \cite{SPT}).

The first comparison of the SZE signal to lensing masses was presented
in \cite{Marrone09} who found a fairly large scatter using a sample of
14 massive clusters. The lensing masses, however, were based on HST
observations covering only the cluster cores. This complicates the
mass estimates and might lead to increased scatter \citep{Hoekstra02}.
More recently, \cite{Marrone11} compared the SZE signal for 18
clusters to weak lensing masses determined by \cite{Okabe10} using
Subaru wide field imaging data. The resulting scaling relation is in
good agreement with earlier studies that assumed hydrostatic
equilibrium. \cite{Marrone11} find that a scatter of $\sim 20\%$ in
weak lensing mass at fixed $Y_{\rm SZ}$, although they note the
scatter depends whether clusters are classified as disturbed or not.

In this section we compare our weak lensing masses to available
measurements of the SZE signal from the literature. We consider two
samples, which partly overlap. \cite{Bonamente08} present results for
38 massive clusters with BIMA and OVRO, 19 of which overlap with our
sample. The second comparison is with the overlap of 18 clusters that
were released in \cite{Planck1} as part of the {\it Planck} early
results. Our aim is not to calibrate the scaling relations but to
examine the potential of the SZE signal as a mass-proxy. The
derivation of a useful scaling relation would require the apertures
for the SZE results to be matched to the lensing radii, as is done for
the X-ray observation in \cite{Mahdavi12}. 

\cite{Bonamente08} list their estimates for $r_{2500}^{\rm SZ}$, based
on a joint analysis of the available SZ and X-ray data. Their results
are reproduced in Table~\ref{tabsz}. As the SZ measurement is a
projected quantity (although not as extreme as the lensing signal), we
compare to the projected lensing masses within an aperture
$r_{2500}^{\rm SZ}$, which are also listed in Table~\ref{tabsz}.
Assuming a constant gas fraction and self-similarity, the SZ signal
$Y$ scales with mass as $M^{5/3}\propto D_{A}^2 E(z)^{-2/3} Y$
\citep[e.g.][]{McCarthy03a,Bonamente08}. We assume this holds and take
$D_{A}^2 E(z)^{-2/3} Y$ as the proxy for mass. The left panel of
Figure~\ref{m_sz} shows the projected lensing mass within
$r_{2500}^{\rm SZ}$ as a function of the SZ signal from
\cite{Bonamente08}.

\begin{table}
\caption{Best fit parameters of the SZE scaling relations \label{parsz}}
\begin{center}
\begin{tabular}{lcccc}
Bonamente     & all   & without A2163 \\
\hline
$\sigma_{\log(Y|M)}$ & $0.10^{+0.05}_{-0.04}$ & $0.06^{+0.05}_{-0.06}$ \\  
$M_0$               & $5.31\pm0.24$        & $5.51\pm0.25$\\
$\alpha$            & $0.48^{+0.08}_{-0.07}$ & $0.57^{+0.09}_{-0.08}$ \\ 
\hline
$\sigma_{\log(M|Y)}$ & $0.05\pm0.02$ & $0.04^{+0.02}_{-0.04}$ \\  
$M_0$               & $5.16\pm0.24$ & $5.43\pm0.25$ \\
$\alpha$            & $0.46\pm0.07$ & $0.55^{+0.09}_{-0.08}$ \\ 
\hline
\hline
{\it Planck}     & all   & without A2163 \\
\hline
$\sigma_{\log(Y|M)}$ & $0.17^{+0.09}_{-0.06}$ & $0.12^{+0.07}_{-0.05}$ \\  
$M_0$               & $11.0\pm0.7$           & $11.2\pm0.7$\\
$\alpha$            & $0.47^{+0.11}_{-0.09}$ & $0.60^{+0.14}_{-0.12}$ \\ 
\hline
$\sigma_{\log(M|Y)}$ & $0.076^{+0.025}_{-0.022}$ & $0.074^{+0.026}_{-0.023}$ \\  
$M_0$               & $10.4\pm0.6$             & $10.6\pm0.6$ \\
$\alpha$            & $0.45\pm0.11$          & $0.56\pm0.16$ \\ 
\hline
\hline
\end{tabular}
\begin{minipage}{\linewidth}
{\footnotesize For the comparison with the measurements from
  \cite{Bonamente08}, $M_0$ is the normalization of the best fit power
  law with slope $\alpha$ to the projected weak lensing aperture mass
  within $r_{2500}^{\rm SZ}$ in units of $10^{14}h_{70}^{-1}$\msun,
  for a pivot $D_{\rm A}^2E(z)^{-2/3}Y=10^{-4}$. We present parameters
  for a log-normal intrinsci scatter in $Y$ for fixed mass,
  $\sigma_{\log(Y|M)}$ and vice versa, where we use the logarithm with
  base 10. For the comparison with the {\it Planck} result from
  \cite{Planck1} we use a pivot of $2\times 10^{-4}$. For both samples
  we present results for the full sample and when Abell~2163 is
  excluded.  }
\end{minipage}
\end{center}
\end{table}

We fit a power law model to the measurements, following the procedure
described in \S5.1. For a pivot value $D_{\rm A}^2E(z)^{-2/3}Y=10^{-4}$
we find a best fit normalization of $M_0=(5.16\pm0.24)\times
10^{14}h_{70}^{-1}$\msun~ and a slope $\alpha=0.48^{+0.08}_{-0.07}$,
somewhat shallower than $\alpha=0.6$ expected from the self-similar
model.  We obtain an intrinsic scatter of $12\pm5\%$ in mass for fixed
$Y$. Note that our scatter is lower, but consistent with the values
found by \cite{Marrone11}. The merging cluster A2163 is the main
outlier and if we remove this cluster the results are consistent with
no intrinsic scatter and the slope steepens to
$\alpha=0.57^{+0.09}_{-0.08}$ in agreement with the self-similar
prediction and the slope of $0.60^{+0.08}_{-0.06}$ found by
\cite{Bonamente08} for their full sample.

\cite{Planck1} provide the integrated $y$-parameter within an aperture
of radius $5\Theta^{X}_{500}$, where $\Theta^{X}_{500}$ is determined
from X-ray observations by \cite{Piffaretti11}. Table~\ref{tabsz}
lists the {\it Planck} measurements and our projected lensing masses
within $\Theta^{X}_{500}$. The right panel of Figure~\ref{m_sz} shows
the weak lensing mass within this radius as a function for the SZE
signal measured by \cite{Planck1}. For these data we choose a pivot
value $D_{\rm A}^2E(z)^{-2/3}Y=2\times 10^{-4}$, which gives a best
fit normalization of $M_{0}=(10.4\pm0.6)\times
10^{14}h_{70}^{-1}$\msun~ and a slope $\alpha=0.45\pm0.11$. The
intrinsic scatter is $19^{+7}_{-6}\%$ in mass for fixed $Y$,
consistent with the result for the \cite{Bonamente08} sample and
\cite{Marrone11}. As before, omitting A2163 from the analysis steepens
the slope to $\alpha=0.56\pm0.16$.

\section{Conclusions}

The Canadian Cluster Comparison Project targeted 50 massive X-ray
luminous clusters of galaxies with redshifts $0.15<z<0.55$, with the
aim of studying the scaling relations between cluster mass and
baryonic tracers and to probe the cluster-to-cluster variation in the
thermal properties of the hot intracluster medium. In this paper we
present the results of our weak lensing analysis of deep wide-field
imaging data obtained using the CFHT.

To ensure a significant detection of the lensing signal, clusters with
an ASCA temperature of $T_X>$5keV from \cite{Horner01} were the main
targets of the CCCP. This led to an overrepresentation of $z\sim 0.2$
clusters. Although the sample lacks a well-defined selection function,
it appears to be representative, based on a comparison of the the
$L_X-T_X$ relation \citep{Mahdavi12}. We include two additional
clusters that were located in the observed field-of-view. Hence we
determine weak lensing masses for a total sample of 52 clusters of
galaxies. We update the masses for 20 clusters studied previously in
\cite{Hoekstra07}, using the mass-concentration relation from
\cite{Duffy08} and present new results for 32 clusters observed with
MegaCam.

We measure the lensing signal out to large radii, which allows us to
determine aperture masses, which are nearly model-independent. To
allow comparison with other observables we deproject the masses. We
also fit NFW models to the data and explore the sensitiviy of our
results to the adopted mass-concentration relation. The values of
$M_{500}$ based on the aperture masses are robust, with a 20\%
increase (decrease) in the normalization of $c(M)$ resulting in a 4\%
increase (6\% decrease) in the mass. The aperture masses agree well
with the results from fitting NFW models. Although we cannot rule out
a lower normalization, a significant increase in the concentration at
a given mass leads to inconsistent values for $M_{500}$ when comparing
the masses from the NFW fit to the aperture masses.

The aperture masses are the reference for the comparison of cluster
properties at other wavelenghts. The scaling relation between a range
of X-ray properties and lensing mass is presented in \cite{Mahdavi12}.
In this paper we limit the comparison to published measurements of the
Sunyaev-Zel'dovich (SZE) effect. We study a sample of 19 clusters that
overlap with the study of \cite{Bonamente08} and another sample of 18
cluster that were observed by {\it Planck} \citep{Planck1}. 

The SZE signal correlated well with the projected lensing mass. For
both samples we find a best fit slope of the power law scaling
relation that is lower than the value of $\alpha=0.6$ for self-similar
models. However, when the merging cluster A2163 is excluded, in both
cases the slopes are in agreement with the self-similar prediction.
We find an intrinsic scatter of $12\pm5\%$ in projected mass
($M_{2500}$) for fixed $Y$ for the clusters that overlap with
\cite{Bonamente08}. The comparison with the {\it Planck} results
\citep{Planck1} yields an intrinsic scatter in projected mass
($M_{500}$) of $19^{+7}_{-6}\%$ at fixed $Y$. 

The scatter agrees well with the results from \cite{Marrone11} and is
comparable to the scatter in the scaling relations with (more
expensive) X-ray observables such as hydrostatic mass and temperature
\citep[e.g.,][]{Mahdavi12, Okabe10b}. This demonstrates that the
SZE signal is a competitive proxy for cluster mass.

\vspace{0.5cm} 

We thank Edo van Uitert for a careful reading of the manuscript. HH
acknowledges support from the Netherlands organisation for Scientific
Research (NWO) through VIDI grant 639.042.814; HH and CB acknowledge
support from Marie Curie IRG Grant 230924. We also acknowledge support
by the National Science and Engineering Research Council (NSERC) and
the Canadian Foundation for Innovation (CFI). This research used the
facilities of the Canadian Astronomy Data Centre operated by the
National Research Council of Canada with the support of the Canadian
Space Agency.

\bibliographystyle{mn2e}
\bibliography{cccp}

\end{document}